\documentclass[preprintnumbers,article,amsmath,amssymb,floatfix,10pt,prd,onecolumn,
superscriptaddress,nofootinbib]{revtex4-2}
\usepackage{bm}
\usepackage{amsfonts}
\usepackage{latexsym}
\usepackage{graphicx}
\usepackage{amsmath,amssymb}
\usepackage[normalem]{ulem}
\usepackage{textcomp}
\usepackage{hyperref}
\hypersetup{colorlinks,citecolor=blue}
\hypersetup{colorlinks=true,linkcolor=red,filecolor=magenta,    urlcolor=blue}
\usepackage{amsmath}
\usepackage{xcolor}
\usepackage{epsfig}

\newcommand{\m}{\mathring}
\newcommand{\be}{\begin{equation}}
\newcommand{\ee}{\end{equation}}
\newcommand{\bea}{\begin{eqnarray}}
\newcommand{\eea}{\end{eqnarray}}
\newcommand{\eeas}{\end{eqnarray*}}
\newcommand{\beas}{\begin{eqnarray*}}

\def\jnl@style{\it}
\def\aaref@jnl#1{{\jnl@style#1}}

\def\aaref@jnl#1{{\jnl@style#1}}

\def\aj{\aaref@jnl{AJ}}                   
\def\apj{\aaref@jnl{ApJ}}                 
\def\apjl{\aaref@jnl{ApJ}}                
\def\apjs{\aaref@jnl{ApJS}}               
\def\apss{\aaref@jnl{Ap\&SS}}             
\def\aap{\aaref@jnl{A\&A}}                
\def\aapr{\aaref@jnl{A\&A~Rev.}}          
\def\aaps{\aaref@jnl{A\&AS}}              
\def\mnras{\aaref@jnl{Mon.~Not.~Roy.~Astron.~Soc.}}             
\def\prd{\aaref@jnl{Phys.~Rev.~D}}        
\def\prc{\aaref@jnl{Phys.~Rev.~C}}  
\def\prl{\aaref@jnl{Phys.~Rev.~Lett.}}    
\def\qjras{\aaref@jnl{QJRAS}}             
\def\skytel{\aaref@jnl{S\&T}}             
\def\ssr{\aaref@jnl{Space~Sci.~Rev.}}     
\def\zap{\aaref@jnl{ZAp}}                 
\def\nat{\aaref@jnl{Nature}}              
\def\aplett{\aaref@jnl{Astrophys.~Lett.}} 
\def\apspr{\aaref@jnl{Astrophys.~Space~Phys.~Res.}} 
\def\physrep{\aaref@jnl{Phys.~Rep.}}      
\def\physscr{\aaref@jnl{Phys.~Scr}}       
\def\commat{\aaref@jnl{Comm.~Math.~Phys.}}              
\def\science{\aaref@jnl{Science}}               
\def\cqg{\aaref@jnl{Classical Quant.~Grav.}}            
\def\jpcs{\aaref@jnl{JPCS}}                                     
\def\ijmpd{\aaref@jnl{Int.~J.~Mod.~Phys.~D}}                    
\def\grg{\aaref@jnl{Gen.~Relat.~Gravit.}}               
\def\rpp{\aaref@jnl{Rep.~Prog.~Phys.}}          
\def\npa{\aaref@jnl{Nucl.~Phys.~A}}        
\def\lrr{\aaref@jnl{Living Rev.~Rel.}}                   
\def\jcap{\aaref@jnl{J.~Cosmology Astropart.~Phys.}}    
\def\rmp{\aaref@jnl{Rev.~Mod.~Phys.}}   
\def\epjc{\aaref@jnl{Eur.~Phys.~J.~C}} 
\def\plb{\aaref@jnl{~Phy.~Lett.~B}} 
\def\mpla{\aaref@jnl{Mod.~Phy.~Lett.~A}} 
\def\arxiv{\aaref@jnl{arxiv.org}}

\allowdisplaybreaks[1]

\begin{document}

\title{Emergent Universe in $f(Q)$ gravity theories}

\author{Hamid Shabani}
\email{h.shabani@phys.usb.ac.ir}
\affiliation{Physics Department, Faculty of Sciences, University of Sistan and Baluchestan, Zahedan, Iran}
\author{Avik De}
\email{avikde@um.edu.my}
\address{Institute of Mathematical Sciences, Faculty of Science, Universiti Malaya, 50603 Kuala Lumpur, Malaysia}
\author{Tee-How Loo}
\email{looth@um.edu.my}
\affiliation{Institute of Mathematical Sciences, Faculty of Science, Universiti Malaya, 50603 Kuala Lumpur, Malaysia}

\footnotetext{The research was supported by the Ministry of Higher Education (MoHE), through the Fundamental Research Grant Scheme (FRGS/1/2023/STG07/UM/02/3, project no.: FP074-2023). }
\begin{abstract}
One resolution of the ancient cosmic singularity, i.e., the Big Bang Singularity (BBS), is to assume an inflationary stage preceded by a long enough static state in which the universe and its physical properties would oscillate around certain equilibrium points. The early period is referred to as the Einstein Static (ES) Universe phase, which characterizes a static phase with positive spatial curvature. A stable Einstein static state can serve as a substitute for BBS, followed by an inflationary period known as the Emergent Scenario. The initial need has not been fulfilled within the context of General Relativity, prompting the investigation of modified theories of gravity. The current research aims to find such a solution within the framework of symmetric teleparallel gravity, specifically in the trendy $f(Q)$ theories. An analysis has been conducted to investigate stable solutions for both positively and negatively curved spatial FRW universes, in the presence of a perfect fluid, by utilizing various torsion-free and curvature-free affine connections. Additionally, we propose a method to facilitate an exit from a stable ES to a subsequent inflationary phase. We demonstrate that $f(Q)$ gravity theories have the ability to accurately depict the emergence of the universe.
\end{abstract}
\maketitle

\section{Introduction}\label{sec1}

One motive for application of modified theories of gravity (MTG) is to resolve the issue of the primordial big bang singularity (BBS) which is accounted for one of the major deficiencies of standard cosmology (SC), together with the absence of a complete theory of quantum gravity (QG). The existence of a singularity within the Planck's scale in which physical quantities such as energy density obtain extreme values is generally distinguished by the verification of curvature invariants in the same way  they diverge at the singularity as well as geodesics extension~\cite{hawking1975}. BBS cannot be fundamentally removed as a characteristic of SC, implying that the classical spacetime has a threshold point beyond which the SC is not applicable. On the other hand, the inflationary scenario during which the Universe experiences an exponential growth in size is widely accepted as the earliest state of the evolution of the Universe and is supported by numerous data resources~\cite{bennett2013,akrami2020}\footnote{See~\cite{achucarro2022} and references therein for a recent review on the current status of inflation focusing the theoretical backgrounds.} is not in accord with BBS~\cite{penrose1989,brandenberger2013}. Hence, different alternatives have been introduced for BBS\footnote{To preserve BBS consistent with the present day observations some attempts had been made e.g., to explain how the observed isotropy of the Universe forms from BBS Misner’s chaotic~\cite{misner1968} and quiescent~\cite{barrow1978} cosmologies put forward.}. We call them under terminology of ``pre-inflationary scenarios" among which we quote quantum cosmologies~\cite{hossain2010,alesci2017}, cyclic scenarios~\cite{khoury2004,Barrow2004}, bouncing cosmologies~\cite{peter2002}, non-local corrections~\cite{bahamonde2017,capozziello2020}, the slow expansion scenario~\cite{piao2003} and emergent universe (EU)~\cite{ellis20041,ellis20042,khodadi,chanda,chanda1}, as the most popular non-singular proposals.

In an emergent structure the universe experiences a past-eternal Einstein static (ES) state\footnote{In mathematical terms it requires the ES phase is stabilized against all types of perturbations.} after which an inflationary evolution is launched. In this way the primitive singularity, or more exactly, the beginning point for time is avoided; the universe ever exists  around a static state (up to small perturbations) for undetermined value of time. Also, EU can be treated classically as the curvature scale of the universe exceeds the Planck scale~\cite{ellis20041,ellis20042}. Beside motivations come from eliminating the BBS there are also some evidences in favor of an unknown physics prior to cosmic inflation; Studies on Cosmic Microwave Background (CMB) data which has been reported by COBE~\cite{smoot1992} and then Planck~\cite{akrami2020} reveal a suppression of the TT-spectrum in CMB which may be explained by pre-inflationary theories.

The ES solution as the basic part of emergent scenario has been studied by Einstein himself in 1917. The original version of the ES universe is discerned as the foremost models of cosmology which consists of FRW metric with positive spatial curvature\footnote{Initially, the positive spatially curvature has been used to guarantee the existence of ES solutions, nevertheless, it was later understood that such a assumption can lead to suppression of the TT mode of CMB~\cite{labrana2015}. It is worth mentioning that examinations of the Planck data agree with a positive curvature~\cite{valentino2019,handley2021}.}, a perfect fluid and a cosmological constant as ingredients~\cite{hawking1975,griffiths2009}. It describes a dynamically static universe since it seemed that the universe does not change in large scales~\cite{raifeartaigh2014}. Despite it really depicts a static universe consistent with that time perceptions, it suffers from instabilities. As a matter of fact, the original ES solution is gravitationally unstable under small homogeneous and isotropic perturbations~\cite{eddington1930} which shows that the universe cannot remain in the ES state for enough long time. Stability of ES solutions has also been investigated under inhomogeneous perturbations in the presence of both pressure-less as well as the ultra-relativistic perfect fluid~\cite{harrison1967}. Nevertheless, it is neutrally stable bearing all types of small perturbations as long as $c_{s}^{2}>1/5$ (here $c_{s}$ denotes the speed of sound) is satisfied~\cite{barrow2003}. Furthermore, it has been shown that the ES solution is not stable against homogeneous perturbations when Bianchi type-IX metric and different types of matter are assumed~\cite{barrow2012}.

As mentioned before, in the earliest stages of the evolution of the universe which is governed under extreme conditions, General Relativity (GR) does not provide a consistent platform to present a correct picture of physical rules. In this regard, in the absence of a comprehensive theory of gravity, a rational choice is modifying GR from different perspectives. For example, it has been shown that adding a squared term of the Ricci scalar to the Einstein-Hilbert action (coming from quantum corrections) leads to an early inflationary de Sitter epoch~\cite{starobinsky1980,starobinsky1983}. The lack of a stable ES solution in the background of GR is another motivation to explore MTG to find admissible solutions which explains EU scenarios; such a proposal must include stable ES solutions which later connect an accelerated expansion stage. Various MTG's has been inspected so far. For instance, we refer to loop quantum gravity~\cite{canonico2010,bag2014}, brane world models~\cite{heydarzade2015,heydarzade2016}, scalar-fluid theories~\cite{boehmer2015}, $f(R)$ gravity~\cite{barrow1983,bohmer2007,seahra2009,sharif20201}, Horava–Lifshitz theory~\cite{bohmer2010,maeda2010}, Einstein-Gauss–Bonnet Gravity~\cite{paul2010,gohain2024}, massive gravity~\cite{paris2012,zhang2013}, metric-Palatini gravity~\cite{boehmer2013}, Einstein–Cartan theory~\cite{atazade2014}, Jordan–Brans–Dicke theory~\cite{huang2014}, $f(R,T)$ gravity ($T$ indicates the trace of EMT)~\cite{shabani2017,sharif20202}, Einstein-Cartan-Brans-Dicke context~\cite{shabani2019}, Rastall theories~\cite{shabani2022}, Energy Momentum Squared Gravity~\cite{khodadi2022,akarsu2023,sharif2023}, scalar-tensor theory~\cite{huang2023},  EU in the background of the modified algebra~\cite{barca2023}, 
the Continuous Spontaneous Localization (CSL) framework~\cite{palermo2022}, Mimetic gravity~\cite{huang2020}, Eddington inspired Born-Infeld theory~\cite{li2017},the Generalised Uncertainty Principle formulation (GUP)~\cite{fadel2022,bosso2022,segreto2023}.

Of late, a popular research trend in MTG is observed, in which torsion and non-metricity, the other two geometric entities of spacetime are delegated to characterise gravity instead of curvature. This particular branch of MTG is termed  teleparallel theory. As a special mention, the modified $f(T)$ gravity ($T$ denotes the torsion scalar), and $f(Q)$ gravity ($Q$ denotes the non-metricity scalar) successfully describe most part of the evolution history of the Universe. For detailed survey on these two theories, look respectively in ~\cite{review_fT} and ~\cite{review_fQ} and the comprehensive references therein. A particular aspect of $f(Q)$ theory not well investigated is its depiction of the Universe by spatially curved FRW model, only limited study in the existing literature is available \cite{FLRW/connection, FLRW/connection1,ad/viability,fQcosmohamid,fQec2,palia,jensko}. We employ the established formulation of $f(Q)$ theory within a spatially curved FRW spacetime, incorporating a variable temporal function $\gamma(t)$. Our present investigation focuses on the EU scenario within this framework.

\section{Symmetric teleparallel formulation}\label{sec2}
In this section we briefly review the geometric framework for the symmetric teleparallel theory of gravity and its generalisation, namely the $f(Q)$ theory of gravity.
In such formulations, in contrast to GR,  a symmetric teleparallel affine connection $\Gamma^\alpha_{\,\,\, \beta\gamma}$ instead of the conventional Levi-Civita connection $\m\Gamma^\alpha_{\,\,\, \beta\gamma}$ of the corresponding metric tensor $g_{\mu\nu}$, is used to define the covariant derivative
as well as to account for the gravitational impact.
The teleparallelism of an affine connection means it is having zero curvature while 
by symmetry it means the corresponding torsion vanishes identically.   
We  define the non-metricity tensor 
corresponding to $\Gamma^\alpha_{\,\,\, \beta\gamma}$ as 
\begin{equation} \label{Q tensor}
Q_{\lambda\mu\nu} = \nabla_\lambda g_{\mu\nu}
:=\partial_\lambda g_{\mu\nu}-\Gamma^\alpha{}_{\mu\lambda}g_{\alpha\nu}
-\Gamma^\alpha{}_{\nu\lambda}g_{\mu\alpha}\,.
\end{equation}
The non-metricity scalar  can hence be defined as 
\begin{equation} \label{Q}
Q= \frac{1}{4}(-Q_{\lambda\mu\nu}Q^{\lambda\mu\nu} + 2Q_{\lambda\mu\nu}Q^{\mu\lambda\nu} +Q_\lambda Q^\lambda -2Q_\lambda \tilde{Q}^\lambda)\,,
\end{equation}
where  
\[
Q_{\lambda}=Q_{\lambda\mu\nu}g^{\mu\nu}; \quad \tilde Q_{\nu}=Q_{\lambda\mu\nu}g^{\lambda\mu}.
\]

As the simplest form of the symmetric teleparallel theory of gravity 
is equivalent to GR, it is natural to introduced  a modified $f(Q)$ gravity \cite{coincident} to extend GR, for which the total action is given as  
\begin{equation*}
S = \frac1{2\kappa}\int f(Q) \sqrt{-g}\,d^4 x
+\int \mathcal{L}_M \sqrt{-g}\,d^4 x\,,
\end{equation*}
where $f(Q)$ is a scalar field.
The dynamics variables for the above system are the metric tensor $g_{\mu\nu}$ and the symmetric teleparallel affine connection $\Gamma^\alpha_{\,\,\, \beta\gamma}$.
Firstly, by varying the preceding action term 
with respect to the metric we obtain the field equation
\cite{zhao}
\begin{equation} \label{FE}
\mathcal E_{\mu\nu} = \kappa T^{(m)}_{\mu\nu}\,,
\end{equation}
where 
\begin{align} \label{FE-01}
T^{(m)}_{\mu\nu}:=&-\frac2{\sqrt{-g}}\frac{\delta(\sqrt{-g}\mathcal L_m)}{g^{\mu\nu}}\,, 
\end{align}
is the energy-momentum (EMT) tensor and 
\begin{align} \label{FE-02}
\mathcal E_{\mu\nu}:=&F \m{G}_{\mu\nu}+\frac{1}{2} g_{\mu\nu} (FQ-f) + 2F' P^\lambda{}_{\mu\nu} \partial_\lambda Q\,.
\end{align}
Here we have defined $F=df/dQ$;
primes denote differentiations with respect to the arguments;
\begin{equation} \label{P}
P^\lambda{}_{\mu\nu} = \frac{1}{4} \left( -2 L^\lambda{}_{\mu\nu} + Q^\lambda g_{\mu\nu} - \tilde{Q}^\lambda g_{\mu\nu} -\frac{1}{2} \delta^\lambda_\mu Q_{\nu} - \frac{1}{2} \delta^\lambda_\nu Q_{\mu} \right) \,,
\end{equation} 
 is  the superpotential tensor 
and
$$\mathring{G}_{\mu\nu} = \mathring{R}_{\mu\nu} - \frac{1}{2} g_{\mu\nu} \mathring{R}\,.$$ 
All the expressions with a $\mathring{()}$ is calculated with respect to the Levi-Civita Connection $\m\Gamma^\alpha_{\,\,\, \beta\gamma}$.
In addition, the EMT tensor $T^{(m)}_{\mu\nu}$ is assumed to be a perfect fluid, which is expressed as 
\begin{align}\label{pf11}
T^{(m)}_{\mu\nu}=(p+\rho)u_\mu u_\nu+pg_{\mu\nu}\,,
\end{align} 
where $\rho$ and $p$ denote the energy density and the  pressure of the ordinary matter.
With the assumption of vanishing hypermomentum tensor
$\Delta_\lambda{}^{\mu\nu}:=-\frac2{\sqrt{-g}}
\frac{\delta(\sqrt{-g}
\mathcal L_m)}{\delta\Gamma^\lambda{}_{\mu\nu}}$
\cite{hyper},
the connection field equation  
\begin{align}\label{FE2}
(\nabla_\mu-L^\alpha{}_{\mu\alpha})(\nabla_\nu-L^\alpha{}_{\nu\alpha})
(F P^{\nu\mu}{}_\lambda)=0\,,
\end{align}
can be obtained after the variation of the action terms
with respect to the affine connection. 
Here $L^\lambda{}_{\mu\nu}$ is the disformation tensor which is  given by
\begin{equation} \label{L}
L^\lambda{}_{\mu\nu} = \frac{1}{2} (Q^\lambda{}_{\mu\nu} - Q_\mu{}^\lambda{}_\nu - Q_\nu{}^\lambda{}_\mu) \,.
\end{equation}
Noticing that 
$\m\nabla^\mu\mathcal E_{\mu\nu}$ is identical with twice the LHS of Eq. (\ref{FE2}), 
the energy conservation is then satisfied:
\cite{de-loo-saridakis}
\begin{align}\label{EC-01}
\kappa\m\nabla^\mu T^{(m)}_{\mu\nu}=2
(\nabla_\mu-L^\alpha{}_{\mu\alpha})(\nabla_\nu-L^\alpha{}_{\nu\alpha})
(F P^{\nu\mu}{}_\lambda)=0\,,
\end{align}



\section{The homogeneous and isotropic model of the universe}\label{sec3}

The background metric for the spatially homogeneous and isotropic spacetime is the FLRW metric whose line element takes the form 
\begin{align}\label{metric}
ds^2 = -\mathrm{d} t^2 
+a\left(t\right)^{2}\left( \frac{dr^2}{1-kr^2} +r^2\mathrm{d}\theta^2+r^2\sin^2\theta\mathrm{d} \phi^2\right)\,, 
\end{align}
where $k=0,\pm1$ denotes the spatial curvature of the Universe.  
Particularly, there is only a unique class of gauge invariant teleparallel symmetric affine connections associated with the metric while $k\neq0$.
It gives rise to a novel insight of the $f(Q)$ dynamics in both spatially flat and spatially curved FLRW background.


The non-trivial connection coefficients are given by \cite{FLRW/connection}
\begin{align}\label{aff}
\Gamma^t{}_{tt}=&-\frac{k+\dot\gamma}\gamma, 
	\quad 					\Gamma^t{}_{rr}=\frac{\gamma}{1-kr^2}, 
	\quad 					\Gamma^t{}_{\theta\theta}=\gamma r^2, 
	\quad						\Gamma^t{}_{\phi\phi}=\gamma r^2\sin^2\theta								\notag\\
\Gamma^r{}_{tr}=&-\frac{k}{\gamma}, 
	\quad  	\Gamma^r{}_{rr}=\frac{kr}{1-kr^2}, 
	\quad		\Gamma^r{}_{\theta\theta}=-(1-kr^2)r, 
	\quad		\Gamma^r{}_{\phi\phi}=-(1-kr^2)r\sin^2\theta,												\notag\\
\Gamma^\theta{}_{t\theta}=&-\frac{k}{\gamma}, 
	\quad		\Gamma^\theta{}_{r\theta}=\frac1r,
	\quad		\Gamma^\theta{}_{\phi\phi}=-\cos\theta\sin\theta,										\notag\\
\Gamma^\phi{}_{t\phi}=&-\frac k\gamma, 
	\quad 	\Gamma^\phi{}_{r\phi}=\frac1r, 
	\quad 	\Gamma^\phi{}_{\theta\phi}=\cot\theta,
\end{align}
where $\gamma(t)$ is any non-zero function of time and 
a dot denote a derivative with respect to time.
It follows from Eq. (\ref{Q}) that the corresponding non-metricity scalar takes the form
\begin{equation}\label{Q-2}
    Q(t)=-3\left[2H^2+\left(\frac{3k}{\gamma}-\frac{\gamma}{a^2}\right)H-\frac{2k}{a^2}-k\frac{\dot{\gamma}}{\gamma^2}-\frac{\dot{\gamma}}{a^2}\right].
\end{equation}
Using the preceding equation and the field equation (\ref{FE}), 
we obtain the modified Friedmann equations
\begin{align}\label{rho}
\rho=&\frac12f+\left(3H^2+3\frac k{a^2}-\frac12Q\right)F+\frac32\dot Q\left(-\frac k\gamma-\frac\gamma{a^2}\right)F'\,,
\end{align}
\begin{align}\label{p}
p=&-\frac12f+\left(-3H^2-2\dot H-\frac k{a^2}+\frac12Q\right)F
        +\dot Q\left(-2H-\frac32\frac k\gamma+\frac12\frac\gamma{a^2}\right)F'\,.
\end{align}
Furthermore, 
as a direct consequence of Eq. (\ref{EC-01}), we obtain the following continuity relation, which can be derived using Eqs. 
(\ref{rho})--(\ref{p}):
  \begin{align}\label{cr}
     \dot{\rho}+3H(p+\rho)=-\frac32\Bigg[\left\{\left(\frac\gamma{a^2}H+3\frac k\gamma H+2\frac{\dot \gamma}{a^2}\right)\dot{Q}
    +\left(\frac{k}{\gamma}+\frac{\gamma}{a^2}\right)\ddot{Q}\right\}F'
    +\left( \frac{k}{\gamma}+\frac{\gamma}{a^2} \right)\dot{Q}^2F''\Bigg]=0.
 \end{align}
\section{ES solutions}\label{sec-}
Here in the present section, as the fist step, we find a solution corresponding to the Einstein static state which may exist when all time derivatives of the engaged variables  do vanish. In fact, In Eqs.~(\ref{Q})--(\ref{cr}) different order of time derivatives of $Q(t)$, $\gamma(t)$, $a(t)$ as well as $H(t)$ must be zero. We apply this discipline to a dynamical system which is defined by $x(t)=a(t)$, $y(t)=\dot{a}(t)$. 
{Motivated by this fact that one preferably may choose the $f(Q)$ function as a model with small deviation from GR~\cite{wang2024}, we use the following function  all throughout the manuscript}
 \begin{align}\label{fq}
f(Q)=\alpha Q+\beta Q^{2}.
 \end{align}
{In the case of the function~(\ref{fq}), setting $\alpha=1$ and $\beta=0$ returns GR, hence, the $Q$ squared dependent term denotes the possible deviations. Also, the implications of the Starobinsky model~\cite{starobinsky1987} motivated us to investigate the present model~\cite{lin2021,araujo2024}. Astronomical data~\cite{wang2024,pradhan2024} put constraints on the free parameters of $f(Q)$ function of type~(\ref{fq}). The models were evaluated under energy condition criteria~\cite{silva2024,fQec2}. In~\cite{nojiri2024}, it was shown that the choice~(\ref{fq}) allows a unification between early-time and late-time accelerated expansions. Besides, the dynamical system studies show that a true sequence of cosmic evolution (a saddle matter dominated era followed by a stable dark energy era) can be achieved in both coincident and non-coincident approaches~\cite{vishwakarma2024, rana2024}}. As one will observe, it will be exhibited that the ES solution can still be studied using such a simple function without loss of generality. In this case for Eqs.~(\ref{rho})--(\ref{cr}) one gets
 \begin{align}\label{rhogamma}
&\rho=\frac{1}{2 \gamma ^4 x^6}\Bigg\{3 \gamma ^2 \bigg[-45 \beta  k^2 x^4 y^2+2 \gamma ^2 x^2 \Big(3 \beta  \left(2 k^2+7 k y^2-6 y^4\right)+\alpha  x^2 \left(k+y^2\right)\Big)\nonumber\\
&+24 \beta  \gamma  k x^3 y \left(k-4 y^2\right)+24 \beta  \gamma ^3 k x y+15 \beta  \gamma ^4 y^2\bigg]-9 \beta  x \left(\gamma ^2+k x^2\right)\bigg[2 \gamma ^2 \dot{y} \Big(\gamma  (\gamma -4 x y)-3 k x^2\Big)\nonumber\\
&+2 \gamma  x \ddot{\gamma} \left(\gamma ^2+k x^2\right)+x \dot{\gamma} \Big(\dot{\gamma} \left(\gamma ^2-3 k x^2\right)-8 \gamma ^2 y^2\Big)\bigg]\Bigg\},
 \end{align}
 \begin{align}\label{pgamma}
&p=\frac{1}{2 \gamma ^4 x^6}\Bigg\{9 \beta  k^2 x^6 \left(5 \dot{\gamma}^2-2 \gamma  \ddot{\gamma}\right)-12 \beta  \gamma ^4 x^3 \Big[\dot{y}(2 \dot{\gamma}+7 k-12 y^2)+2 y \ddot{\gamma}\Big]+\nonumber\\
&6 \beta  \gamma ^5 x \left(8 y^3+\gamma \dot{y}+2 y \dot{\gamma}\right)-9 \beta  \gamma ^6 y^2+\gamma ^2 x^4 \bigg[27 \beta  k^2 y^2-2 \alpha  \gamma ^2 \left(k+y^2\right)\nonumber\\
&+6 \beta  k \Big(-2 \gamma  \ddot{\gamma}+\dot{\gamma}^2+4 \dot{\gamma} \left(k-5 y^2\right)+36 \gamma  y \dot{y}\Big)\bigg]\nonumber\\
&3 \beta  \gamma ^4 x^2 \bigg[2 \left(\gamma  \ddot{\gamma}+2 k^2+7 k y^2-6 y^4\right)+\dot{\gamma} \left(3 \dot{\gamma}+8 k-8 y^2\right)-24 \gamma  y \dot{y}\bigg]\nonumber\\
&-2 \gamma  x^5 \bigg[\gamma  \dot{y} \Big(2 \alpha  \gamma ^2+3 \beta  k \left(4 \dot{\gamma}-9 k\right)\Big)+6 \beta  k y \left(2 \gamma  \ddot{\gamma}-4 \dot{\gamma}^2+9 k \dot{\gamma}\right)\bigg]\Bigg\},
 \end{align}
and 
 \begin{align}
&\dot{\rho}+3\frac{y}{x}(p+\rho)=\Sigma\label{cemt}\\
&\Sigma=-\frac{9 \beta }{\gamma ^5 x^7}\Bigg\{\gamma ^3 y^2 \big(3 k^2 x^4 y+8 \gamma ^3 x \left(k-y^2\right)+9 \gamma ^4 y\big)+x\bigg[-4 \gamma ^4 x^2 \dot{y}^2 \left(\gamma ^2+k x^2\right)-\nonumber\\
&2 k x^4 \dot{\gamma}^3 \left(\gamma ^2+3 k x^2\right)-2 \gamma  x y \dot{\gamma}^2 \left(\gamma ^4+6 k^2 x^4+\gamma ^2 k x^2\right)+2\gamma ^2\dot{y}\Big[-2 \gamma ^2 k x^3 \left(k-2 y^2\right)\nonumber\\
&+x \dot{\gamma} \left(\gamma ^4+3 k^2 x^4-4 \gamma ^3 x y\right)-2 \gamma ^4 x \left(k-4 y^2\right)-4 \gamma ^5 y\Big]+\gamma \dot{\gamma}\Big[\gamma  y \big(3 k^2 x^4 y+8 \gamma ^3 x \left(y^2-k\right)-7 \gamma ^4 y\big)\nonumber\\
&+2 x^2 \ddot{\gamma} \left(\gamma ^4-3 k^2 x^4-2 \gamma ^2 k x^2\right)\Big]\gamma ^2 +x \left(\gamma ^2+k x^2\right) \Big[\dddot{\gamma} x \left(\gamma ^2+k x^2\right)+\gamma  \ddot{y} \left(\gamma ^2-3 k x^2-4 \gamma  x y\right)\nonumber\\
&+2 y \ddot{\gamma} \left(3 k x^2-\gamma ^2\right)\Big]\bigg]\Bigg\},\nonumber
 \end{align}

where all functions' arguments have been dropped for the sake of clarity. Thereafter, the Einstein static solution can be obtained by setting all time derivatives equal to zero. This gives  

\begin{align}
\rho_{ES}=\frac{3 k \left(6 \beta  k+\alpha  x_{ES}^2\right)}{x_{ES}^4},\label{rhoES}\\
p_{ES}=\frac{k \left(6 \beta  k-\alpha  x_{ES}^2\right)}{x_{ES}^4}\label{pES},
 \end{align}

and Eq.~(\ref{cemt}) is automatically satisfied (both sides vanish). By assuming that a perfect fluid with $p=w\rho$ fills the Universe, we obtain the radius of the Einstein static state as

\begin{align}\label{aES}
a_{ES}=\sqrt{6 k \frac{1-3 w}{1 +3w}\frac{\beta }{\alpha}}.
 \end{align}

Consequently, the values of equilibrium quantities are determined by both physical properties of matter, the equation of state (EoS) parameter $w$, the underlying gravity model (via model parameters $\alpha$ and $\beta$) and the spatial curvature, $k$. In the following sections, we exploring different choices of $\gamma(t)$ we seek for a true stability of the solution~(\ref{aES}). In all our discussions, setting $x_{ES}=a_{ES}=1$ simplifies the analysis. This choice not only results in certain key equations becoming independent of the constants $\alpha$ and $\beta$, but also leads to a more concise and manageable form. The condition $x_{ES}=1$ gives

\begin{align}\label{alpha}
\alpha= 6 k \frac{1-3 w}{1 +3w}\beta.
 \end{align}

\section{Case: $\gamma(t)=\epsilon a(t)$ with $\epsilon=\pm 1$ and $k=-1$}\label{sec-m1}
In \cite{ad/viability}, it was shown that in the case of $\gamma(t)=\pm a(t)$ for $k=-1$ the EMT tensor is conserved in $f(Q)$ gravity. In fact, in this case the $\Sigma$ expression in Eq.~(\ref{cemt}) vanishes. Hence, we use Eqs.~(\ref{rho})--(\ref{p}) to obtain possible oscillating solution. In this case, assuming a perfect fluid, the dynamical system equivalent of Eqs.~(\ref{rho})--(\ref{p}) in terms of $x$ and $y$ read 

\begin{align}
&\dot{x}=y\label{ds1-1},\\
&\dot{y}=-\frac{\left(9 w^2-1\right) (y-1) \Big[(y-1)^2 (3 y+1)-x^2 (y+1)\Big]}{2 x \Big[-3 w \left(x^2-6 (y-1)^2\right)+x^2+6 (y-1)^2\Big]}.\label{ds1-2}
\end{align}
 
One obtains $\dot{y}=0$ for $x=1$ and $y=0$ which are the required conditions to have the solution~(\ref{aES}) when $k=-1$ is set. To consider the stability properties of the solution~(\ref{aES}) the eigenvalues of the Jacobian matrix of the system~(\ref{ds1-1})--(\ref{ds1-2}) must be analyzed, which are 

\begin{align}\label{eig1}
\lambda_{1,2}=\pm i\sqrt{\frac{(3w+1)(3w-1)}{15 w+7}}.
\end{align}

\begin{figure}[ht!]
\begin{center}
\epsfig{figure=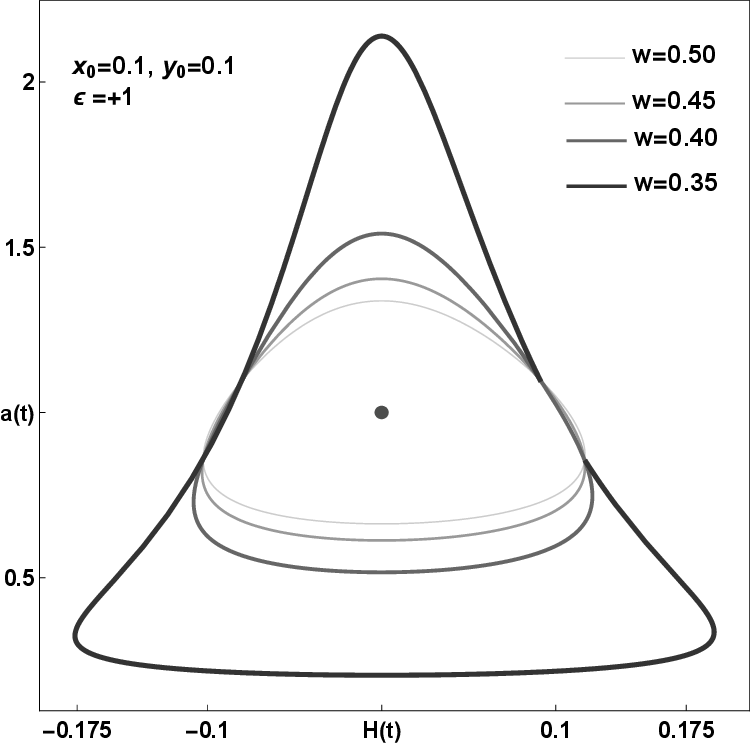,width=7.3cm}\hspace{5.5mm}
\epsfig{figure=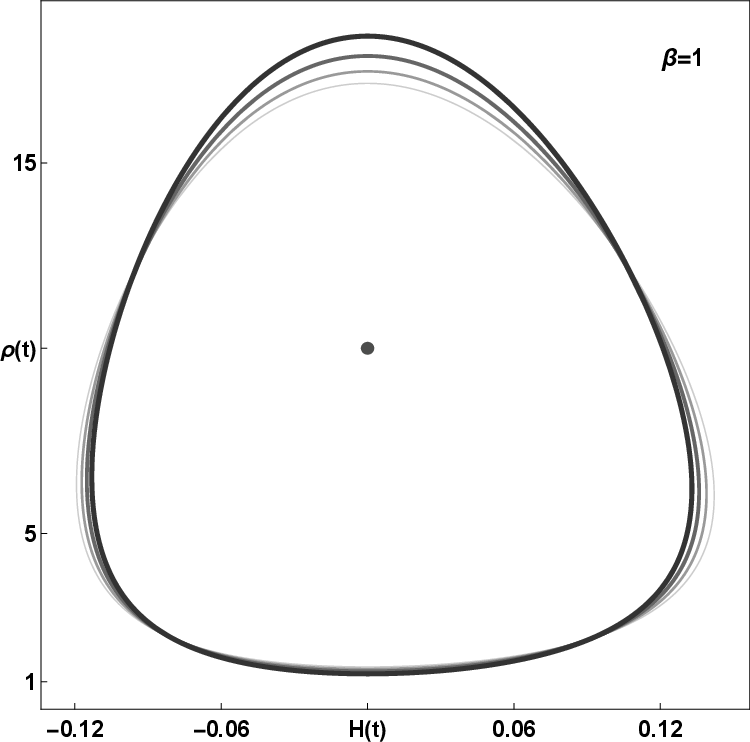,width=7.3cm}\vspace{2mm}
\epsfig{figure=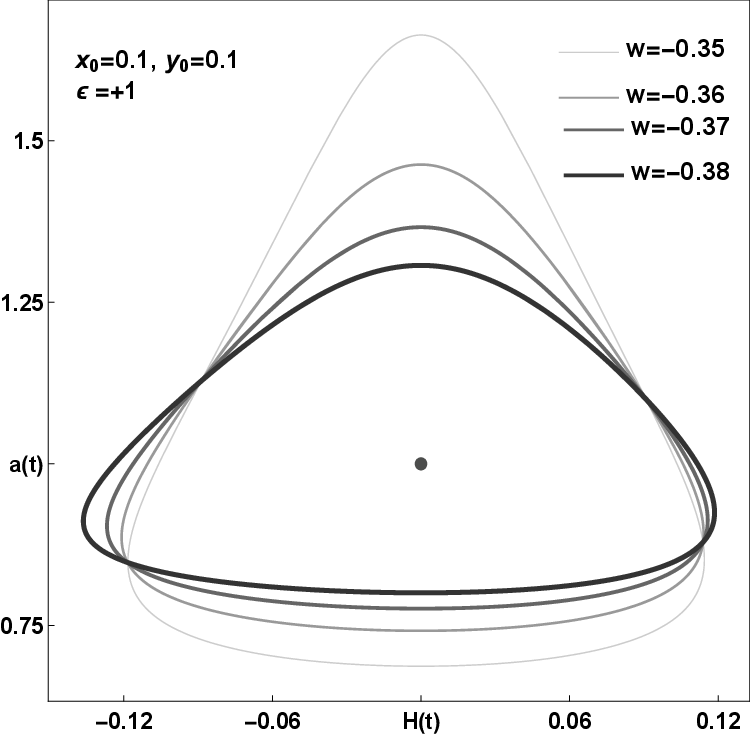,width=7.3cm}\hspace{5.5mm}
\epsfig{figure=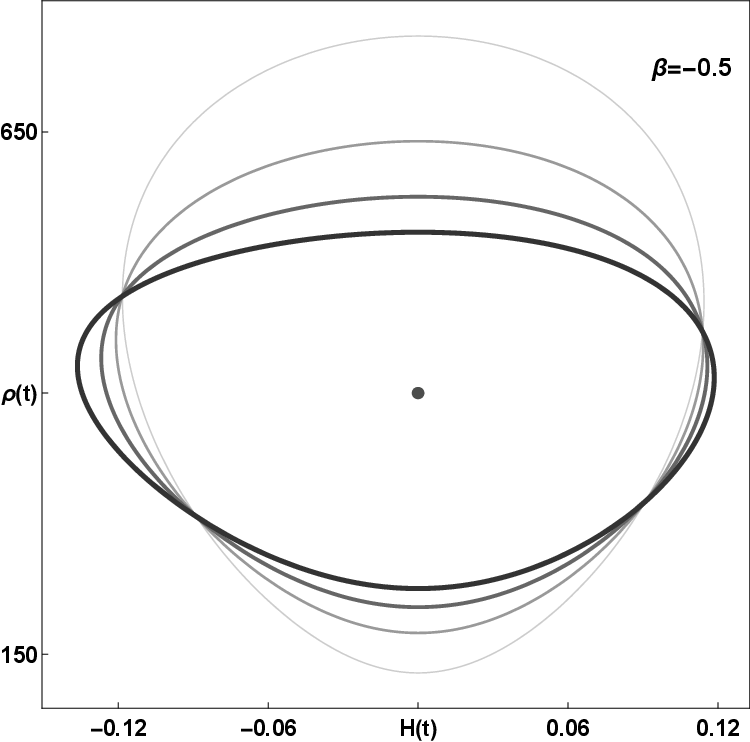,width=7.3cm}
\caption{Different cosmological quantities depicted for $\gamma(t)=a(t)$ and $k=-1$. Left panels: closed $H-a$ curves on which the values which have been used to draw each row are displayed. Right panels: $H-\rho$ plots. The initial values used to solve 
Eqs. (\ref{ds1-1})-(\ref{ds1-2}) are represented by $x_0$ and $y_0$.}\label{m1f1}
\end{center}
\end{figure}

It is interesting that the eigenvalues~(\ref{eig1}) would not include the real-valued parts if the expression inside the square root sign gets positive values. This gives the ranges $-\frac{7}{15}<w<-\frac{1}{3}\land w>\frac{1}{3}$. As a result, the gravitational model under the selection~(\ref{fq}) accepts oscillatory solutions for $\gamma(t)=\pm a(t)$ in a spatially open curved spacetime if a perfect fluid with the mentioned intervals of the EoS parameter defines the cosmic matter. An illustration of engaged quantities for both sides of valid values of $w$ is presented in Fig.~\ref{m1f1}.

\section{Case: $\gamma(t)=\gamma_0$}\label{sec-m2}
Technically, we can use different forms of $\lambda$ functions for which the continuity equation~(\ref{cemt}) is not conserved except at the coordinate ($x=1, y=0$) corresponding to the solution~$a_{ES}=1$. Our strategy in these situations is to obtain $\dot{y}$ from Eqs.~(\ref{rhogamma})--(\ref{pgamma}) for $p=w\rho$ and choose appropriate values of the constants to have minimal variations of the expression~(\ref{cemt}). Note that this method is not irrational because of the nature of oscillatory solutions. It means that the behavior of the expression~(\ref{cemt}) is as small amplitude oscillations around zero value. Thus, in the case of $\gamma(t)=\gamma_0$ we have the following system. 

\begin{align}
&\dot{x}=y,\label{ds2-1}\\
&\dot{y}=\frac{\mathcal{N}}{\mathcal{D}},\label{ds2-2}\\
&\mathcal{N}=9 k^2 (3 w+1) (5 w+1) x^4 y^2+\lambda _0\Bigg\{-24 k w (3 w+1) \left(k-4 y^2\right)x^3 y \nonumber\\
&+\lambda _0 \bigg[2 \left(9 w^2-1\right) \Big(2 k^2 \left(x^2-1\right)+k \left(2 x^2-7\right) y^2+6 y^4\Big) x^2-\lambda _0 (3 w+1)  \Big(8 x \left(3 k w-2 y^2\right)+3 \lambda _0 (5 w y+y)\Big)y\bigg]\Bigg\}\nonumber\\
&\mathcal{D}=2\Bigg\{9 k^2 (w-1) (3 w+1) x^5+x\lambda _0\bigg[12 k (w-3) (3 w+1) x^3 y+\nonumber\\
&\lambda _0 \Big(2 x^2 \big(k \left(3 w \left(3 w-2 x^2+8\right)+2 x^2+7\right)-12 (3 w+1) y^2\big)+\lambda _0 (3 w+1) \big(12 (w+1) x y-\lambda _0 (3 w+1)\big)\Big)\bigg]\Bigg\}.\nonumber
\end{align}

To investigate possible oscillatory solutions, we look for conditions for which the real parts of the eigenvalues of the Jacobian matrix of the system~(\ref{ds2-1})--(\ref{ds2-2}) vanish. Actually, this happens for the following values of the parameters.

\begin{align}
&w=0,~~~~~~~\lambda_{1,2}=\pm i\frac{2 k \gamma _0}{\sqrt{-9 k^2+18 k \gamma _0^2-\gamma _0^4}},~~~k=+1,\label{eig2.1}\\
&\gamma _0=\pm 1,~~~~~\lambda_{1,2}=\pm i\sqrt{\frac{(3w+1)(3w-1)}{15 w+7}},~~~~k=-1.\label{eig2.2}
\end{align}

\begin{figure}[ht!]
\begin{center}
\epsfig{figure=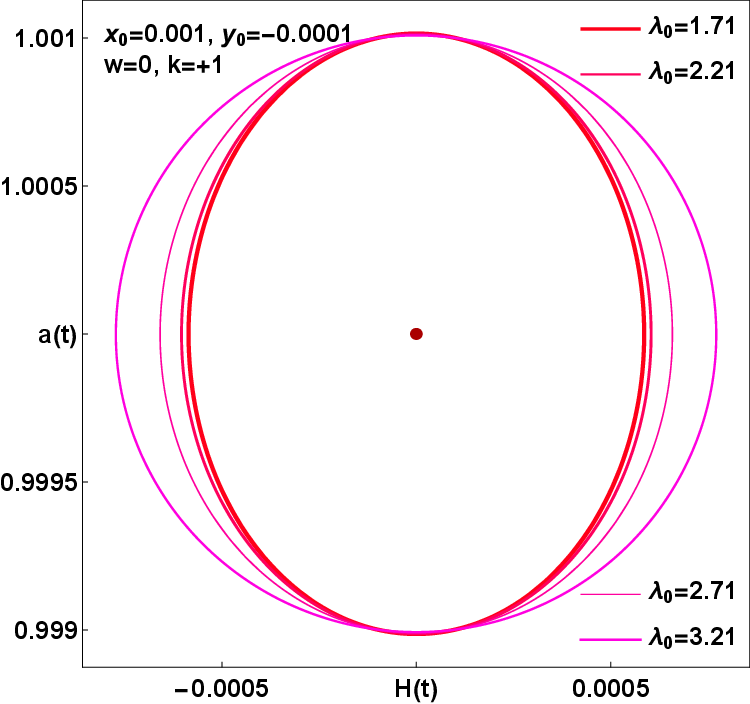,width=7.3cm}\hspace{3.7mm}
\epsfig{figure=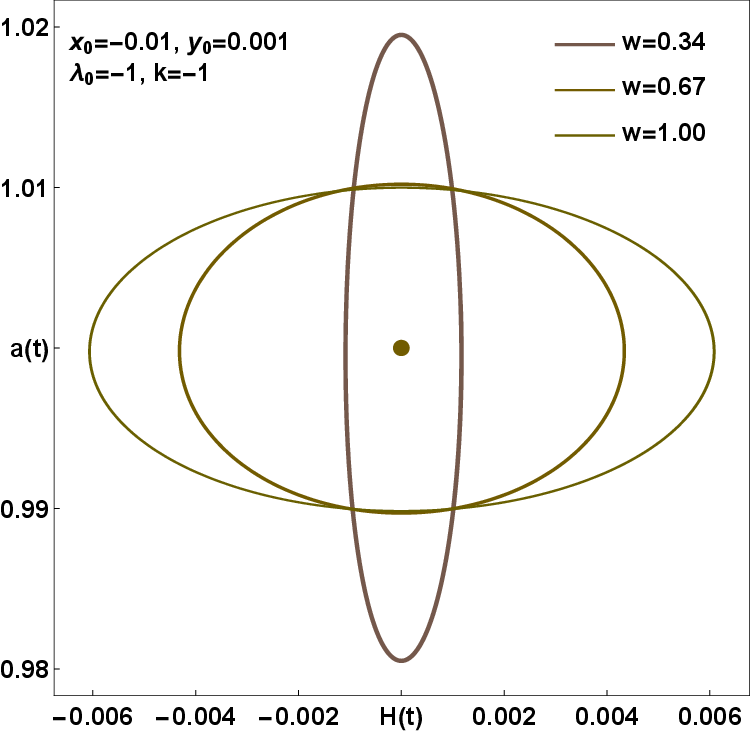,width=7.3cm}\vspace{2mm}
\epsfig{figure=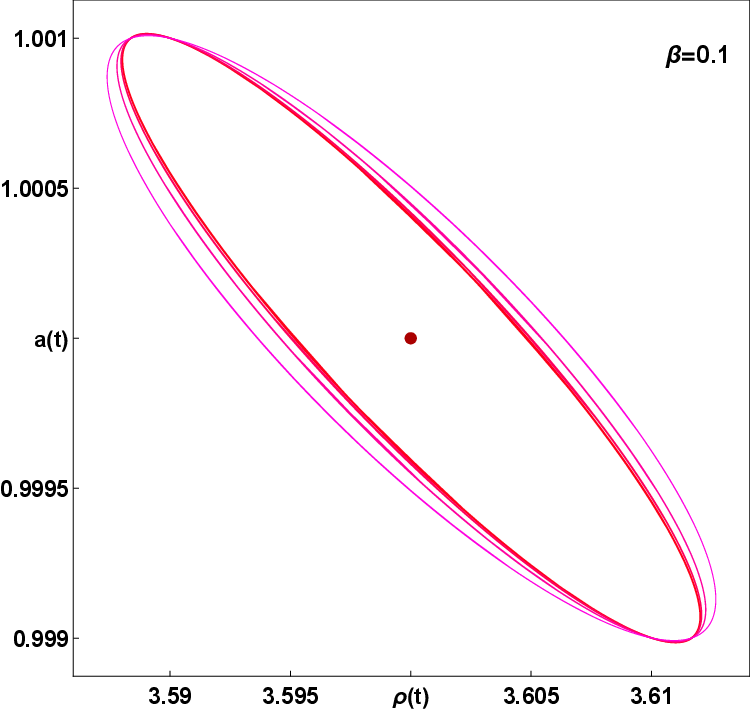,width=7.3cm}\hspace{3.7mm}
\epsfig{figure=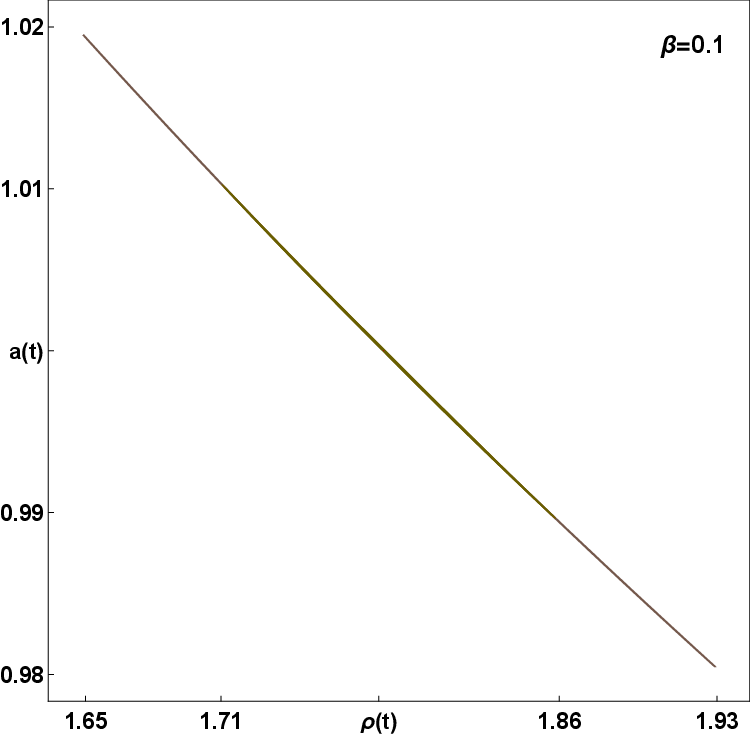,width=7.3cm}\vspace{2mm}
\epsfig{figure=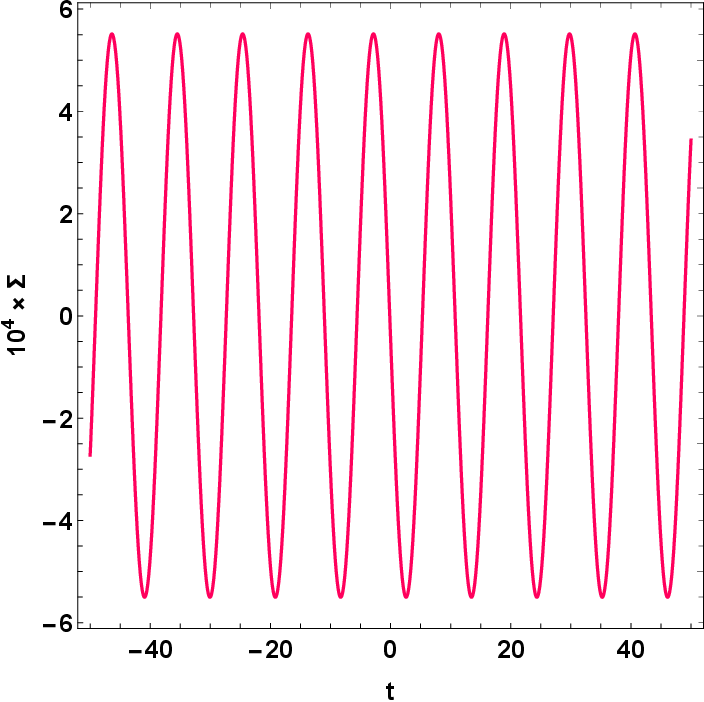,width=7.3cm}\hspace{3.7mm}
\epsfig{figure=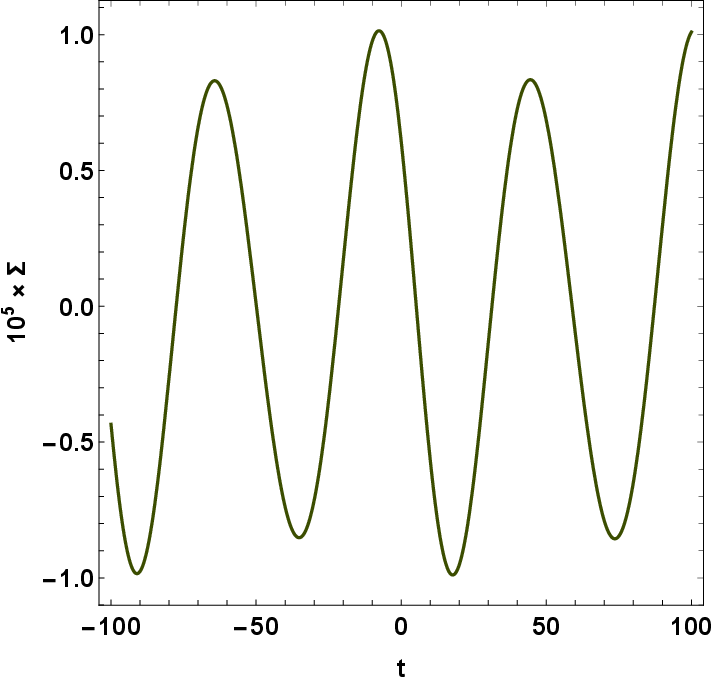,width=7.3cm}
\caption{Cosmological quantities for models with $\gamma=\gamma_0$. Lower left panel has been drawn for $\lambda_0=1.71$ and also for the lower right one we chosen $w=0.34$. The left and the right bottoms have been plotted for $k=+1$ and $k=-1$, respectively. }\label{m2f1}
\end{center}
\end{figure}

Solutions~(\ref{eig2.1}) and~(\ref{eig2.2}) exist for both a closed and an open spatial curvature universe, respectively. The former exists for approximately $-4.18<\gamma _0<-0.71~\land~ 0.71<\gamma _0<4.18$ and the latter holds for again $-\frac{7}{15}<w<-\frac{1}{3}\land w>\frac{1}{3}$. Thus, models with $\gamma(t)=\gamma_0$ represent oscillatory behaviors for some intervals of $w$ and $\lambda_0$. As Fig.~\ref{m2f1} shows, the function $\Sigma$ varies around zero value with amplitude about $10^{-4}$ and $10^{-5}$.

\section{Case: $\gamma(t)=\gamma_0 a^{n}(t)$}\label{sec-m3}
As outlined in Sect.~\ref{sec-m2}, the system of equations in the case of $\gamma(t)=\gamma_0 a^{n}(t)$ reads

\begin{align}
&\dot{x}=y\label{ds3-1}\\
&\dot{y}=\frac{\mathcal{N}}{\mathcal{D}}.\label{ds3-2}
\end{align}
\begin{align}
&\mathcal{N}=2 (3 w-1) \bigg[ \Big(k (7-(n-2) n)+2 x^2\Big)y^2+2 k \left(k+x^2\right)-6 y^4\bigg]x^2 \nonumber\\
&+\gamma_{0}^{-2}y x^{-2 n}\bigg[3 k^2 (n-3) \Big(n (w-3)+5 w+1\Big) x^4 y+8 \gamma _0^3  \Big(y^2 \left(n^2+3 n w-2\right)-k (n-3 w)\Big)x^{3 n+1}\nonumber\\
&-8 \gamma _0 k (n-3 w) \Big(k+(n-4) y^2\Big)x^{n+3} -\gamma _0^4(n+1)  (9 n w+5 n-15 w-3) x^{4 n}y\bigg],\nonumber\\
&\mathcal{D}=2x\Bigg\{(1+3w)^{-1}\bigg[-2 x^2 \Big(k \big(-9 (n-1) w^2+n+24 w+7\big)+(6 w-2) x^2-12 (3 w+1) y^2\Big)\bigg],\nonumber\\
&+3\gamma _0^{-2} k^2 (n-3) (w-1) x^{2(2- n)}-4 k \gamma ^{-1}_{0} (2 n+3 w-9) x^{3-n}y\nonumber\\
&-4 \gamma _0  (2 n+3 w+3) x^{n+1}y\gamma _0^2 +(n+1) (3 w+1) x^{2 n}\Bigg\}.\nonumber
\end{align}

The eigenvalues of system~(\ref{ds3-1})--(\ref{ds3-2}) reveals that only for $n=3w$ one may obtain pure imaginary values. In this case the eigenvalues get the simpler form

\begin{align}\label{eig3}
\lambda_{1,2}=\pm 2  \gamma _0 k i\sqrt{\frac{9 w^2-1}{9 k^2 (w-1)^2 (3 w+1)+6 \gamma _0^2 k \Big[ \left(9 w^2-3 w-7\right)w-3\Big]+\gamma _0^4 (3 w+1)^3}}.
\end{align}

 Consequently, oscillatory behaviors take place when $n=3w$ is set, provided the expression under the root square sign in~(\ref{eig3}) becomes positive. Besides, in the case of $n=3w$ we have

\begin{align}
&\Sigma=-9 \beta  \gamma _0^{-2} x^{-6 w-7}\Bigg\{3 k^2 (w-1)  \bigg[\left(9 w^2-1\right) y^3-9 w x y \dot{y}+x^2 \dddot{y}\bigg]x^4+2 k \gamma _0^2 (3 w-1)  \Big(3 w y \dot{y}+x \dddot{y}\Big)x^{6 w+3}\nonumber\\
&\gamma _0^4 +(3 w+1)  \Big((15 w-8) x y \dot{y}+9 (w-1) (3 w-1) y^3+x^2 \dddot{y}\Big)x^{12 w}-4 k \gamma _0  \Big(\dot{y} \left(k+x \dot{y}-2 y^2\right)+x y \dddot{y}\Big)x^{3 w+4}\nonumber\\
&-4 \gamma _0^3  \bigg[\Big(\left(k+2 (3 w-2) y^2+x \dot{y}\right)+x y \dddot{y}\Big)\dot{y} x+2 (3 w-1) y^2 \left(k-y^2\right)\bigg]x^{9 w+1}\Bigg\}.
\end{align}

\begin{figure}[ht!]
\begin{center}
\epsfig{figure=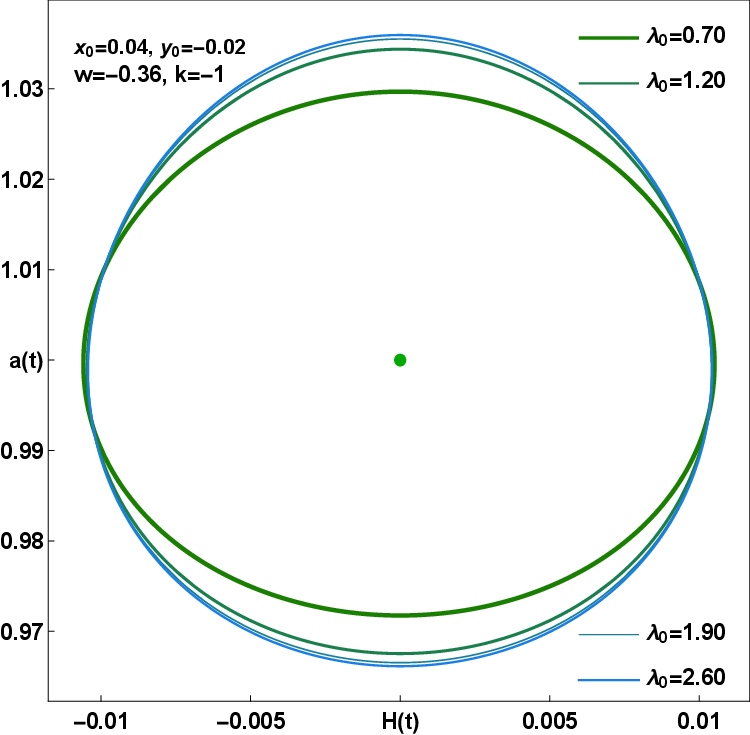,width=7.3cm}\hspace{3.7mm}
\epsfig{figure=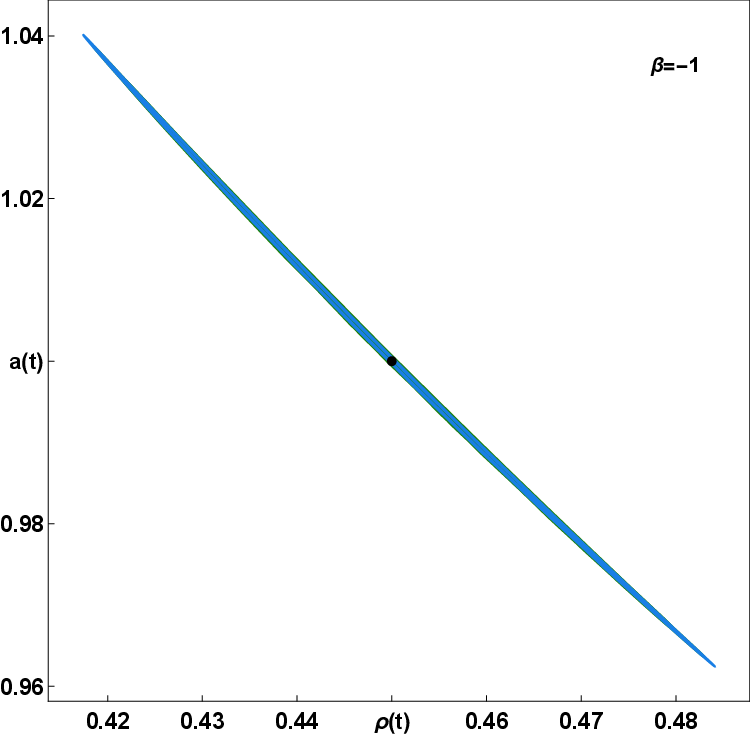,width=7.3cm}\vspace{2mm}
\epsfig{figure=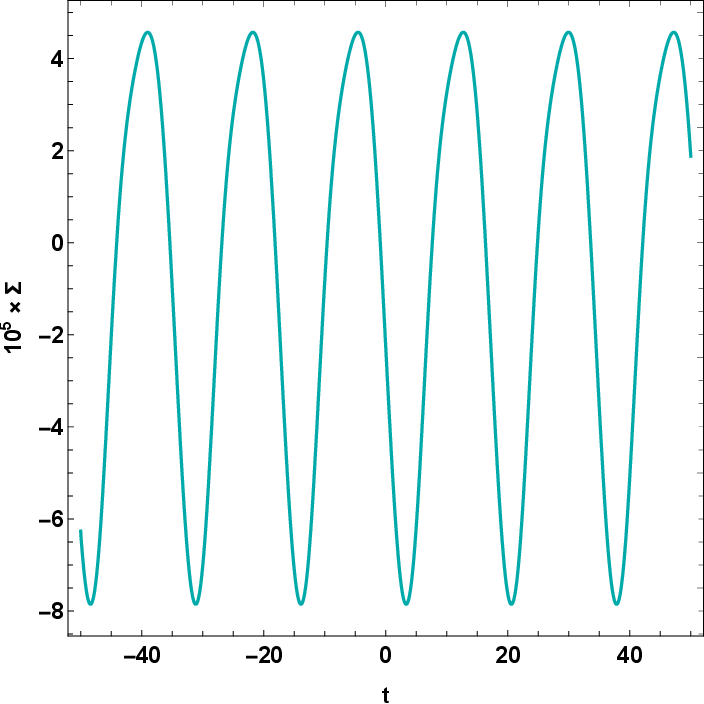,width=7.3cm}\hspace{3.7mm}
\epsfig{figure=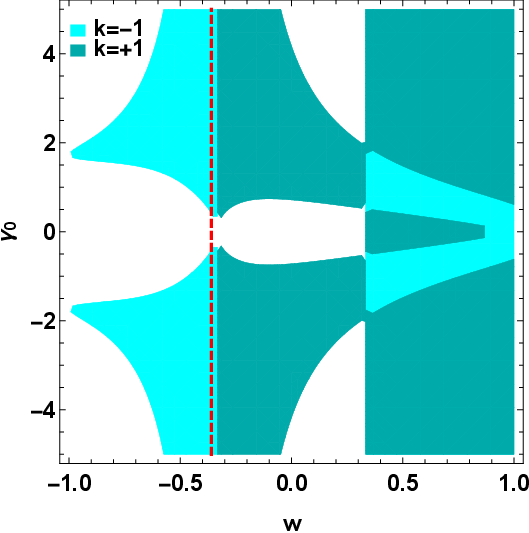,width=7.3cm}
\caption{Properties of the oscillatory dynamics when $\gamma(t)=\gamma_0 a^{n}(t)$ with $n=3w$ is selected. Lower right panel demonstrates $(w,\gamma_0)$ space of oscillatory solution. The red dashed line in the region plot shows the line of $w=-0.36$, for which other panels in this figure have been provided.}\label{m3f1}
\end{center}
\end{figure}

In Fig.~\ref{m3f1} we have plotted an example of oscillatory dynamics for $w=-0.36$ (indicated by a red dashed line) which corresponds to $n=-1.08$ and other values of the model parameters which are denoted on the upper left panel. Again, all the diagrams are provided for small variation of $\Sigma$ expression around zero. A region plot of $w$ and $\gamma_0$ for purely imaginary eigenvalues~(\ref{eig3}) has been provided in lower right panel of Fig.~\ref{m3f1}.

\section{Case: General temporal function $\gamma(t)$}
\label{sec-m4}

Considering the Einstein static solution without any prior restriction on $\gamma(t)$ may be the most exciting case. The corresponding dynamical system is explored in this section.

\begin{align}
&\dot{x}=y\label{ds4-1}\\
&\dot{y}=\frac{\mathcal{N}}{\mathcal{D}}.\label{ds4-2}\\
&\mathcal{N}=(3 w+1)\Bigg\{-4 \gamma ^3 k  \bigg[(1-3 w)xg+6 w\left(k-4 y^2\right)y\bigg]x^3-3  (5 w+1)\gamma ^6 y^2+3 k^2 (5-3 w) x^6 \dot{\gamma}^2\nonumber\\
&+2 \bigg[(3 w+1)gx-12 k w y+8 y^3+2 y \dot{\gamma}\bigg]\gamma ^5 x
\gamma ^2 +k\bigg[\Big(9 k (5 w+1) y-8 g x\Big)y+8 \Big(k-(3 w+5) y^2\Big)\dot{\gamma} \nonumber\\
&+2(1-3 w) \dot{\gamma}^2\bigg]  x^4 +2 k\bigg[3  k (w-1)g x+2  \left(4 \dot{\gamma}-9 k\right)y \dot{\gamma}\bigg]\gamma  x^5+\bigg[-8 g x y \nonumber\\
&+2 (3 w-1) \Big(2 k^2 (x^2-1)+k (2 x^2-7 y^2+6) y^4\Big)+ \Big(8 k+3 (w+1) \dot{\gamma}-8 (3 w+1) y^2\Big)\dot{\gamma}\bigg]\gamma ^4 x^2\Bigg\} \nonumber\\
&\mathcal{D}=2 \gamma ^2 x\Bigg\{9 k^2 (w-1) (3 w+1) x^4+2 k \bigg[ \Big(3 w (3 w-2 x^2+8)+2 x^2+7\Big)\gamma\nonumber\\
& +6 (w-3) (3 w+1) x y\bigg]\gamma x^2+(3 w+1) \bigg[- (3 w+1)\gamma ^2+12  (w+1)\gamma  x y-24 x^2 y^2\bigg]\gamma ^2 +4 (3 w+1)\left(\gamma ^2+k x^2\right) x^2 \dot{\gamma}\Bigg\}\nonumber.
\end{align}

The eigenvalues of the system~(\ref{ds4-1})--(\ref{ds4-2}) cannot be calculated easily. Instead, we discuss the real part of the eigenvalues of the Jacobian matrix $J_{ij}=\partial \dot{q}/\partial q,~q=x,y $ which are

\begin{align}\label{eigJ}
\lambda_{1,2}=\frac{1}{2} \left[\pm\sqrt{\left(J_{11}-J_{22}\right){}^2+4 J_{12} J_{21}}+J_{11}+J_{22}\right].
\end{align}
Therefore, pure imaginary values demand $J_{11}+J_{22}=0$ and $\left(J_{11}-J_{22}\right){}^2+4 J_{12} J_{21}<0$. The former gives a second-order differential equation which determines $\gamma(t)$ for two constants of integration. The equation is derived as

\begin{align}\label{g}
&\ddot{\gamma}=g,\\
&g=\frac{\mathcal{M}}{\mathcal{E}}\nonumber\\
&\mathcal{M}=-6  k w \left(\gamma ^2+k\right) \bigg[9 k^2 (w-1) (3 w+1)+18  k (w+1)^2\gamma ^2+ -(3 w+1)^2\gamma ^4\bigg]\gamma ^2-\nonumber\\
& \Bigg\{81 k^4 (w-1) (3 w+1)+6  k^3 \bigg[w (51 w+26)+15\bigg]\gamma ^2+12 k^2 (3 w+1) (5 w-4)\gamma ^4 \nonumber\\
&+6 k \bigg[7 w (3 w+2)+1\bigg]\gamma ^6 +(3 w+1)^2\gamma ^8 \Bigg\}\dot{\gamma }+\Bigg\{9 k^3 (w-1) (3 w-7) (3 w+1)\nonumber\\
&+3  k^2 \bigg[w \left(45 w^2-39 w-41\right)+3\bigg]\gamma ^2+3  k (3 w+1) \bigg[3 w (w+2)+7\bigg]\gamma ^4+ -(3 w+1)^2(3 w+5)\gamma ^6\Bigg\}\dot{\gamma }^2\nonumber\\
&+16 k (3 w+1)  \left(\gamma ^2+k\right)\dot{\gamma}^3.\nonumber\\
&\mathcal{E}=2\gamma (\gamma ^2+k)\bigg[9 k^2 (w-2) (w-1) (3 w+1)+6 k w \left(9 w^2-6 w-7\right) \gamma ^2\nonumber\\
&+(3 w+2)(3 w+1)^2\gamma ^4 +4 (3 w+1) \left(\gamma ^2+k\right)\dot{\gamma}\bigg].\nonumber
\end{align}

\begin{figure}[ht!]
\begin{center}
\epsfig{figure=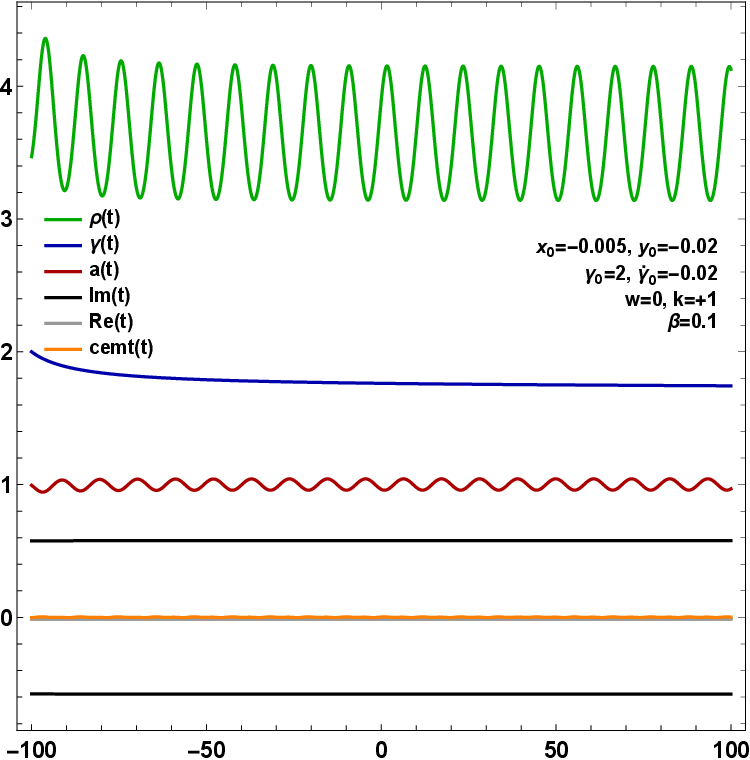,width=9cm}
\caption{Important quantities numerically obtained for the model $f(Q)=\alpha Q+\beta Q^{2}$ with $\alpha= 6 k \frac{1-3 w}{1 +3w}\beta$. Here, $\gamma(t)$ function is obtained by solving eq.~(\ref{g}). Also, $Im(t)$ and $Re(t)$ are the real and imaginary parts of~(\ref{eigJ}).}\label{m4f1}
\end{center}
\end{figure}

The latter condition is also guarantied for solution of Eq.~(\ref{g}) and some suitable model parameters $w$, $k$ and initial values for $x$ and $y$, as well\footnote{Note that the model parameters $\alpha$ and $\beta$ do not appear in equations~(\ref{ds4-2}) and~(\ref{g}).}, provided a small amplitudes for the $\Sigma$

\begin{align}\label{s4}
&\Sigma=-\frac{9 \beta }{\gamma ^5 x^7}\Bigg\{3 k^2\bigg[-2 \gamma  g y+\gamma ^2 \ddot{y}-2 \dot{\gamma } \gamma  \dot{y}+4 \dot{\gamma }^2 y\bigg]\gamma x^6-\bigg[-8 k y+8 y^3+7 \dot{\gamma } y+8 \gamma  \dot{y}\bigg]\gamma ^6 x y +9 \gamma ^7 y^3+\nonumber\\
&k\bigg[4 \gamma  g y+3 k y^3-2 \Big(2 \gamma  \dot{y} \left(k-2 y^2\right)+\gamma ^2 \ddot{y}+\dot{\gamma }^2 y\Big)\bigg]\gamma ^3 x^4 +\bigg[-2 \Big(\gamma  g+\dot{\gamma } \left(\dot{\gamma }+4 k-4 y^2\right)\Big)y\nonumber\\
&+2 \left(\dot{\gamma }-2 k+8 y^2\right) \gamma  \dot{y}+\gamma ^2 \ddot{y}\bigg]\gamma ^5 x^2 + \bigg[\gamma  \dddot{\gamma }+2 \dot{\gamma } g-4 \left(\gamma  y \ddot{y}+\gamma  \dot{y}^2+2 \dot{\gamma } y \dot{y}\right)\bigg]\gamma ^5 x^3 +k^2 \bigg[\gamma ^2 \dddot{\gamma }+6 \dot{\gamma }^3-6 \gamma  \dot{\gamma } g\bigg] x^7\nonumber\\
&k\bigg[2 \gamma ^2 \dddot{\gamma }+\left(2 \dot{\gamma }^2-4 \gamma  g+3 k y^2\right)\dot{\gamma } -4 \gamma ^2 y \ddot{y}-4 \gamma ^2 \dot{y}^2\bigg]\gamma ^2 x^5 \Bigg\}
\end{align}

is attained. Consequently, to find oscillatory equation, one must (numerically) solve equations~(\ref{ds4-1})--(\ref{ds4-2}) and (\ref{g}), simultaneously. Figure.~\ref{m4f1} displays an example of the oscillatory solution numerically extracted for the system~(\ref{ds4-1})--(\ref{ds4-2}) and (\ref{g}). In this figure the orange curve shows the $\Sigma$ expression~(\ref{s4}) with amplitude $[-0.005,0.005]$ around zero, the blue curve depicts $\gamma(t)$ for which the equation~(\ref{g}) is satisfied. The real and imaginary parts of~(\ref{eigJ}) have been indicated by gray (overlapping with the $\Sigma$ plot) and black curves, respectively. The green and red plots show the matter density and the scale factor, respectively. 
{The initial values denoted on Fig.~\ref{m4f1} have been chosen so as to satisfy a few conditions: i) the eigenvalues~(\ref{eigJ}) must be pure imaginary valued to guaranty oscillatory behavior in $x$ and $y$, simultaneously, to lead to a real-valued $\gamma$ function\footnote{This demand makes the affine connection coefficients~(\ref{aff}) real-valued, as well.} which is obtained from Eq.~(\ref{g}). ii) the $\Sigma$ function~(\ref{s4}) (which indicates deviation of the continuity equation (see the first line in Eq.~(\ref{cemt}))) gets small amplitude giving rise small deviations.}
%
%
\section{From the state of Einstein static to the emergent universe}\label{sec-ex}
In this section we attempt to answer the classic question - can the Universe exit from an oscillatory state under $f(Q)$ theory and what is the next possible evolutionary stage. Amidst varied possibilities, we only consider the models studied in Sects.~\ref{sec-m1} and~\ref{sec-m3}. We have hitherto presumed a perfect fluid with constant EoS parameter and obtained conditions for stable ES solutions for which all physical quantities oscillate around their equilibrium centers. Nothing is in favor of an escape from a stable ES state. To allow such a possibility, one may inspect a running EoS parameter whose values leave those regions of $w$ that permit a stable ES solutions. This situation is not unreasonable, given that the physical processes responsible for deviations from equilibrium remain poorly understood. Factors such as dark energy production or quantum effects may alter the properties of cosmic matter, potentially contributing to such deviations.

\begin{figure}[ht!]
\begin{center}
\epsfig{figure=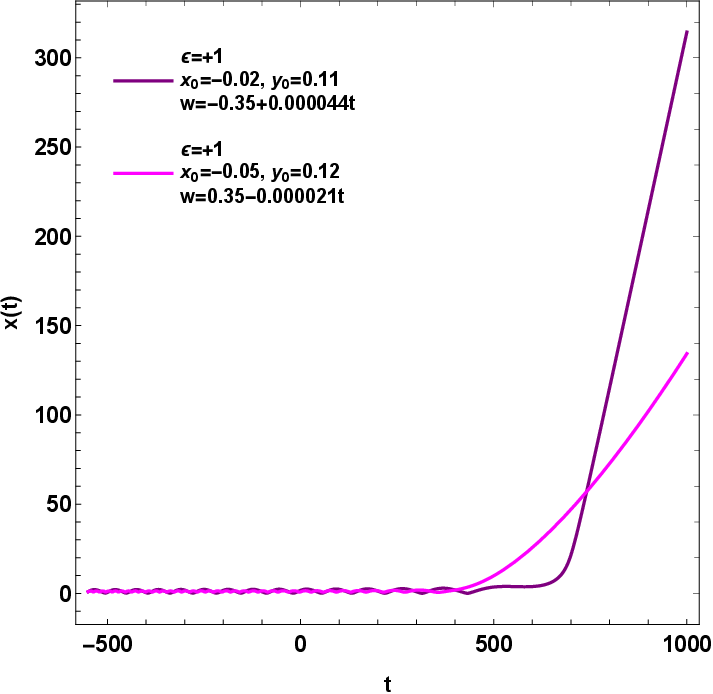,width=8cm}\hspace{2mm}
\epsfig{figure=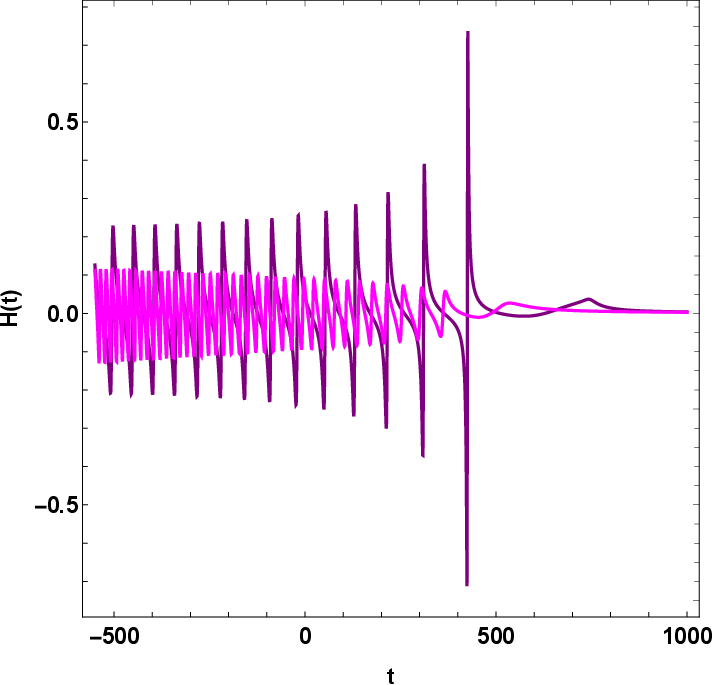,width=8cm}\vspace{2mm}
\epsfig{figure=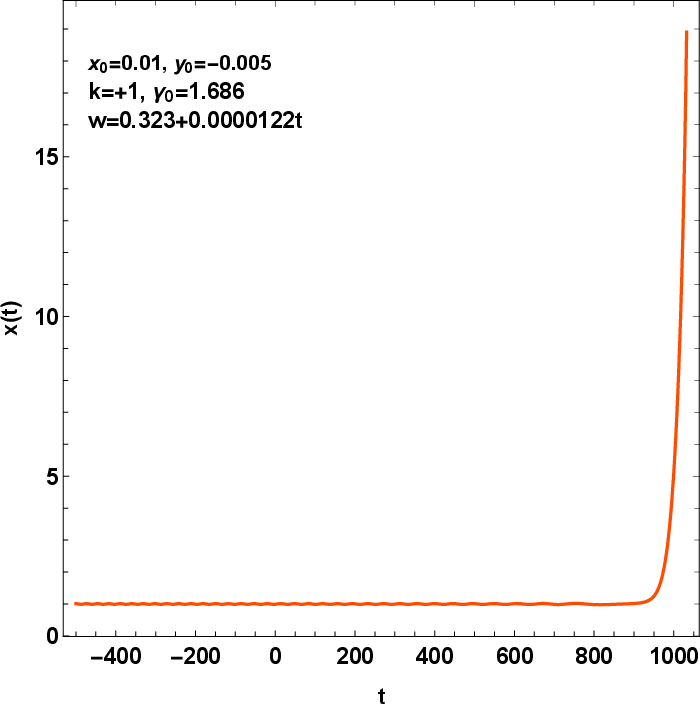,width=8cm}\hspace{2mm}
\epsfig{figure=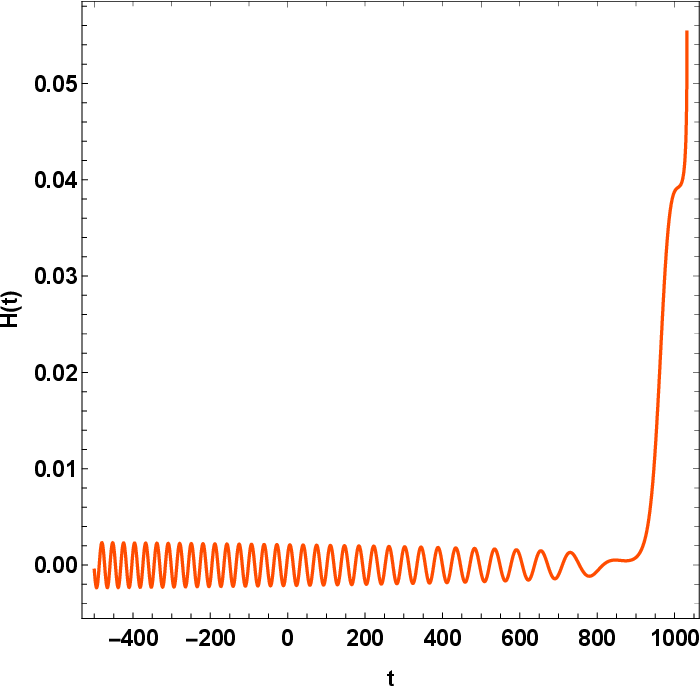,width=8cm}
\caption{Transition from an oscillatory to a decelerated expansion eras which is implied by the model with $\gamma(t)= a(t)$ with $k=-1$ in the upper panels and $\gamma(t)=\gamma_0 a^{n}(t)$ in the lower panels. The former illustrates a decelerating phase after the ES stage while the latter includes an accelerated expansion.}\label{exit}
\end{center}
\end{figure}

As discussed in Sect.~\ref{sec-m1}, in case of $\gamma(t)=\pm a(t)$ with $k=-1$, one has seen that the Universe evolves in an oscillatory way for $-\frac{7}{15}<w<-\frac{1}{3}~\land~w>\frac{1}{3}$. Once $w$ dynamically falls out of these ranges, the eigenvalues of the system~(\ref{ds1-1})--(\ref{ds1-2}) become purely real valued. In this case, the oscillatory pattern in the evolution of quantities disappears. In fact, the solution $(a_{ES}=1, H_{ES}=0)$ would be unstable in such a case. Numerical solutions for linearly varying functions of $w$, which permit transitions beyond the specified intervals for $w$, reveal that the system ~(\ref{ds1-1})--(\ref{ds1-2}) accepts the solution $a(t)=t$. This implies that the Universe experiences a decelerated expansion after an oscillatory phase. Fig.~\ref{exit} demonstrates the phase transition. The upper right panel of Fig.~\ref{exit} depicts the behavior of the Hubble parameter which in this case decreases to zero. The behavior of the Universe under such description can be justified by considering this fact that ($y=1, x=a$) is another solution of Eqs.~(\ref{ds1-1})--(\ref{ds1-2}).

\begin{figure}[ht!]
\begin{center}
\epsfig{figure=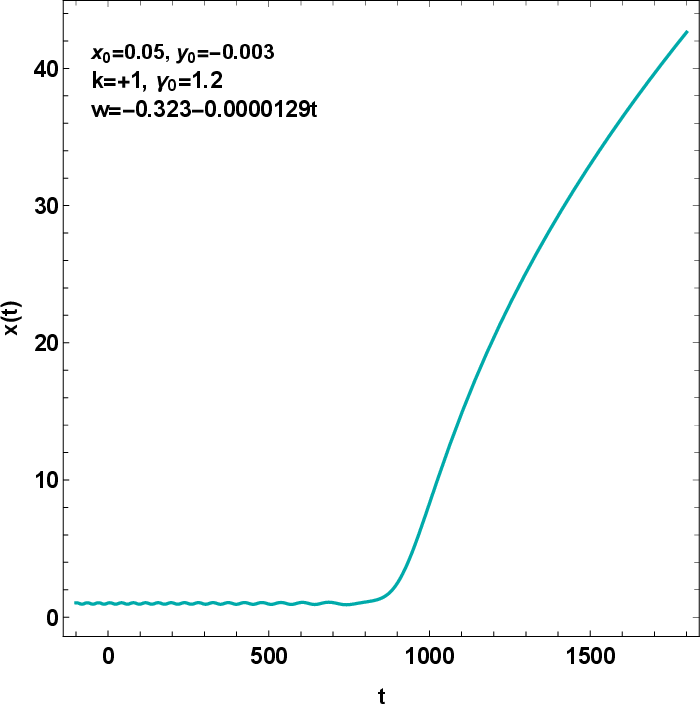,width=8cm}\hspace{2mm}
\epsfig{figure=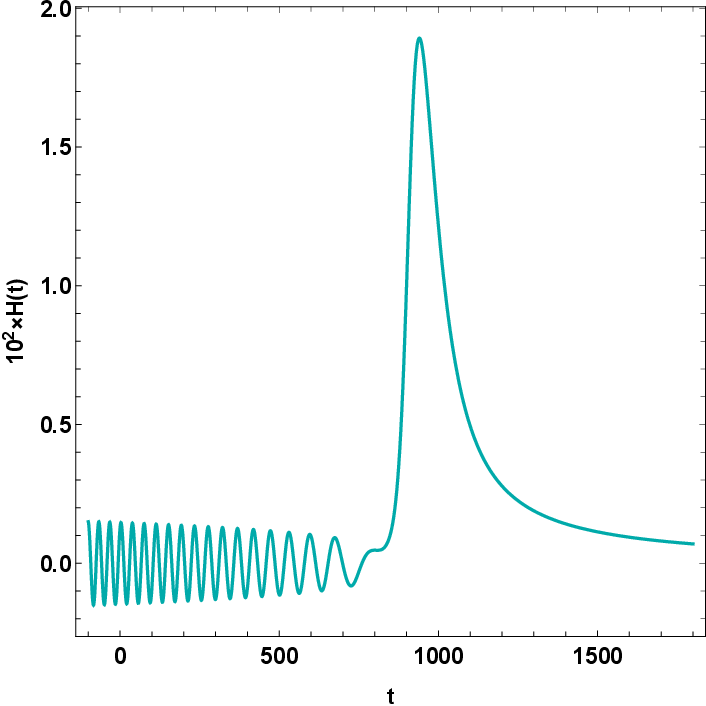,width=8cm}\vspace{2mm}
\epsfig{figure=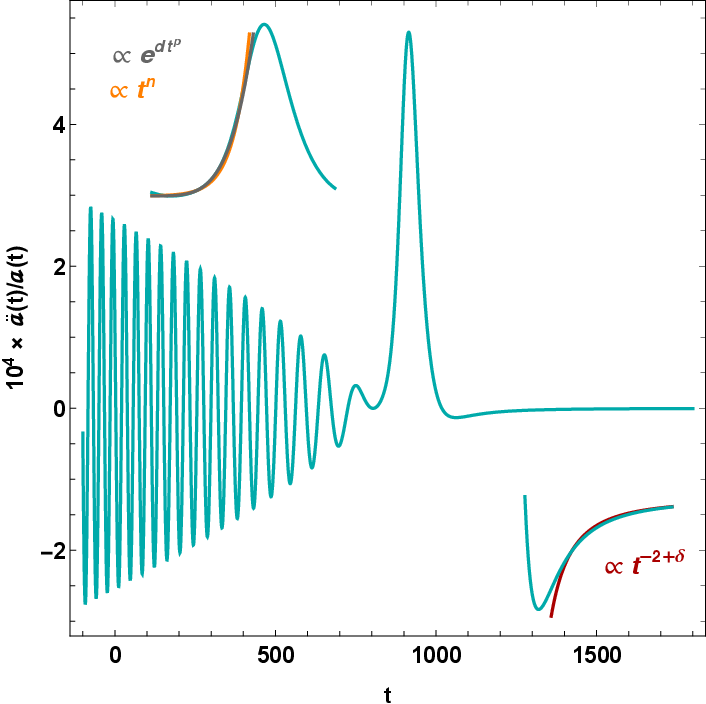,width=8cm}
\caption{A decreasing form of the EoS parameter under consideration of the model $\gamma(t)=\gamma_0 a^{n}(t)$ exhibits a sequence of phase transitions; past-eternal ES state to a temporal accelerated expansion which ends with a decelerated expansion. The last two phases behaves as $e^{dt^{p}}$ (or $t^{n}$) and $t^{-2+\delta}$ for small values of $\delta$, respectively.}\label{exitt}
\end{center}
\end{figure}

Nevertheless, in case of $k=+1$, an EU solution can be achieved by choosing suitable parameters. An example is represented in the lower panels of Fig.~\ref{exit}, which has been extracted under the model discussed in~\ref{sec-m3}. Crossing the value $w=1/3$ to larger values (see the region plot in Fig.~\ref{m3f1}) an inflationary era may begin. In this case the Hubble parameter increases in a non-linear way which confirms an onset of inflationary era. 
{This mentioned accelerating behavior employs an increasing function of the EoS parameter, however, as authors of Ref.~\cite{barrow2003} discusses, a decreasing function may be more reasonable since the ES state should decay to a de Sitter phase (with $w_{ds}=-1$) after which the standard hot big-bang takes over~\cite{ellis20042}. \cite{ellis20042} presents a simple mechanism to happen the process of the transition from the ES state to the de Sitter phase; a (decelerated) minimally coupled scalar field with a slowly varying potential may destabilize the ES state. Based on such an explanation, one may consider that the EoS parameter may be affected by action of a minimally coupled scalar field to enter the invalid regions\footnote{One can define the density and the pressure of a perfect fluid as the ones of a single minimally coupled scalar field. Hence, a running EoS parameter is defined as the ratio of the pressure to the density of the scalar field (see Appendix A in~\cite{ellis20042}). }. By this motivation, we have plotted $\ddot{a}/a$ for a decreasing function of $w$ in Fig.~\ref{exitt}. In this figure, the EoS parameter decreases from a value of slightly greater than $-1/3$  to smaller values. Interestingly, three distinct phases can be distinguished; a past eternal ES state which is decaying to an accelerated expansion period ($\ddot{a}/a\propto(\exp(dt^{p})~\land~t^{n})$ for constants $d$, $p$ and $n$) with a decelerated expansion phase as the final stage ($\ddot{a}/a\propto t^{-2+\delta}$). The decelerated expansion phase may imply initiation of the standard hot big-bang. }

\section{Concluding remarks}\label{sec8}

Recent astronomical data strongly suggest that the physics of an early universe is consistent with an early accelerated expansion mechanism for which the field of Inflaton is responsible. Such an evolutionary process allows solving of unknown phenomena that already pertain to BBS. For example, the horizon problem, the flatness problem and the magnetic monopole as well\footnote{For some recent works see~\cite{lakhal2019,singal2024}.}. However, the inflationary paradigm is not flawless; it cannot follow the BBS since pre-inflationary theories have been invented to introduce a singularity-free beginning point for the inflationary era. A past-eternal stage of the ES period solves the issue. Thereafter, the universe being subjected to an abrupt accelerated expansion. The whole process is called the Emergent Scenario.

In this paper, we show that the emergent scenario which consists of a stable ES solution along with a successful exit to inflationary regime, can also be studied in the background of $f(Q)$ theories. $f(Q)$ models have the capacity to clarify phenomena in both early-time and late-time cosmology without requiring the incorporation of dark energy, the inflaton field, or dark matter. The importance of $f(Q)$ theories in addressing cosmological tensions in recent times is particularly significant. These theories give interesting potential for altering our knowledge of gravity and its manifestations in cosmology. As a particular case the most non-trivial $f(Q)$ model, $f(Q)=\alpha Q+\beta Q^{2}.$ has been studied in our present work. 

In due process, we have dealt with an arbitrary temporal function $\gamma(t)$ that appears in the connection components~(\ref{aff}) as a free parameter. First, we have assumed a few different forms of the function $\gamma(t)$ to obtain the corresponding stable ES solutions. As a second approach, we have used Eqs.~(\ref{rho})--(\ref{p}) to constrain $\gamma(t)$ that permits stable oscillatory solutions to appear. In the former approach, three types of $\gamma(t)$ functions have been investigated;

\begin{itemize}
\item[I)] the case $\gamma(t)=\pm a(t)$ with $k=-1$ for which the conservation of EMT is guarantied in $f(Q)$ gravity. In such cases, a stable ES solutions exist for $-\frac{7}{15}<w<-\frac{1}{3}\land w>\frac{1}{3}$.

\item[II)] models with $\gamma(t)=\gamma_0$ (where $\gamma_0$ is a free constant) are also studied. The ES solution is obtained in two classes: i) $w=0$ with approximately $-4.18<\gamma _0<-0.71~\land~ 0.71<\gamma _0<4.18$ for a close spatial curvature and\\
ii) $\gamma _0=\pm 1$ with $-\frac{7}{15}<w<-\frac{1}{3}\land w>\frac{1}{3}$ for an open spatial curvature.

\item[III)] in the case of $\gamma(t)=\gamma_0 a^{n}(t)$ stable ES solutions can be obtained for $n=3w$. In this case, there exist stable solutions, provided a particular regions in the $(w,\gamma_0)$ plane for $k=\pm 1$ is used to make sure that the expression inside the root square sign~(\ref{eig3}) is positive. See the lower right panel in Fig.~\ref{m3f1}. As can be seen, stable ES solutions are gained when one uses a power-law form of $\gamma$ in both closed and open spatially curved geometry.
\end{itemize}

For the final approach, instead of imposing any additional assumption on $\gamma(t)$

\begin{itemize}
\item[IV)] we finally extracted the differential equation~(\ref{g}) from the equations of motion~(\ref{ds4-1})--(\ref{ds4-2}), through which a function $\gamma(t)$ is obtained. The result has been enlightened in Fig.~\ref{m4f1}.
\end{itemize}

It should also be mentioned that in the case of the item I the EMT is conserved, unlike the items II-IV. In the latter case our solutions ensure the expression $\Sigma$ in $\dot{\rho}+3H(p+\rho)=\Sigma$ to shows oscillations with small amplitude around zero (see the related plots in Figs.~\ref{m2f1}--\ref{m4f1}).

As the next step, we have briefly discussed a process in which the Universe can exit from the past eternal ES era and subsequently fall into accelerated expansion regime. 
Our solutions have been obtained by assuming that there is a perfect fluid with constant EoS parameter which interact with gravitation. 
Regarding such premise the Universe experiences an ever-existing oscillatory state and consequently the desired exit does not happen. 
Nevertheless, if the EoS parameter smoothly runs to invalid regions (those that leads to the non-existence of ES solutions), interesting results can be achieved. 
{Such a process can be occurred by evolution of a single minimally coupled scalar field with a slowly varying potential}. As an example, we have explored the models which are listed as items I and III. 
We have used a linear time-varying forms of EoS with suitable coefficients so that they can evade intervals of $w$ mentioned within the list. 
{Figs.~\ref{exit} and~\ref{exitt} indicate the results. Under the former case the Universe experiences a decelerated expansion in which the scale factor is proportional to time (see the upper panels in Fig.~\ref{exit}). 
On the contrary, the latter models lead the Universe to go into an accelerated expansion regime when the spatial curvature is positive. The lower panels in Fig.~\ref{exit} provide an exit for a EoS parameter which increases to values beyond $1/3$ while when the EoS parameter decreases from the marginal values of $-1/3$ successive phase transitions from a past-eternal ES state to a final decelerating one with an accelerated behavior in the middle era can be achieved, as Fig.~\ref{exitt} illustrates.}




\begin{thebibliography}{90}
\bibitem{hawking1975} S. W. Hawking and G. F. R. Ellis, \textit{The large scale structure of Space-Time}, Cambridge (1975).
\bibitem{bennett2013} C. L. Bennett et al., \textit{Astrophys. J.} \textbf{208}, 2 (2013).
\bibitem{akrami2020} Y. Akrami et al. (Planck), \textit{Astron. Astrophys.} \textbf{641}, A10 (2020).
\bibitem{achucarro2022} A. Ach\'{u}carro et al., arxiv: 2203.08128 [astro-ph.CO].
\bibitem{penrose1989} R. Penrose, \textit{Annals N. Y. Acad. Sci.} \textbf{571}, 249 (1989).
\bibitem{brandenberger2013} R. H. Brandenberger, 
\textit{Lect. Notes Phys.} \textbf{863}, 333 (2013) [arXiv:1203.6698 [astro-ph.CO]].
\bibitem{misner1968}  C. W. Misner, \textit{Astrophys. J.} \textbf{151}, 431 (1968).
\bibitem{barrow1978}  J. D. Barrow, \textit{Nature} \textbf{272}, 211 (1978).
\bibitem{hossain2010} G. M. Hossain, V. Husain and S. S. Seahra, \textit{Phys. Rev. D} \textbf{81}, 024005 (2010).
\bibitem{alesci2017} E. Alesci   et. al.,
\textit{Phys. Rev. D} \textbf{96}, 046008 (2017).
\bibitem{khoury2004}  J. Khoury, P. J. Steinhardt and N. Turok, \textit{Phys. Rev. Lett.} \textbf{92}, 031302 (2004) [hep-th/0307132].
\bibitem{Barrow2004}  J. D. Barrow, D. Kimberly and J. Magueijo, \textit{Class. Quant. Grav.} \textbf{21}, 4289 (2004) [astro-ph/0406369].
\bibitem{peter2002} P. Peter and N. Pinto-Neto, \textit{Phys. Rev. D} \textbf{66}, 063509 (2002).
\bibitem{bahamonde2017}S. Bahamonde, S. Capozziello and K. F. Dialektopoulos, \textit{Eur. Phys. J. C} \textbf{77}, 722 (2017).
\bibitem{capozziello2020} S. Capozziello, M. Capriolo and S. Nojiri, \textit{Phys. Lett. B} \textbf{810}, 135821 (2020).
\bibitem{piao2003} Y. Piao, E. Zhou, \textit{Phys. Rev. D} \textbf{68}, 083515 (2003).
\bibitem{ellis20041} G. F. R. Ellis and R. Maartens, \textit{Class. Quant. Grav.} \textbf{21}, 223 (2004) [gr-qc/0211082].
\bibitem{ellis20042} G. F. R. Ellis, J. Murugan and C. G. Tsagas, \textit{Class. Quant. Grav.} \textbf{21}, 233 (2004) [gr-qc/0307112].
\bibitem{khodadi}M. Khodadi, \textit{Phys. Dark Univ.} \textbf{36} 101013 (2022).
\bibitem{chanda}A. Chanda et. al., \textit{Eur. Phys. J. C}, \textbf{84} 658 (2024).
\bibitem{chanda1}A. Chanda et. al., \textit{Eur. Phys. J. C}, \textbf{84} 658, (2024).


\bibitem{smoot1992} G.F. Smoot et. al., \textit{Astrophys. J.} \textbf{396}, L1 (1992).
\bibitem{labrana2015} P. Labrana, \textit{Phys. Rev. D} \textbf{91}, 083534 (2015).
\bibitem{valentino2019} E. Di Valentino, A. Melchiorri, and J. Silk, \textit{Nature Astronomy} \textbf{4}, 196 (2019).
\bibitem{handley2021} W. Handley, \textit{Phys. Rev. D} \textbf{103}, L041301 (2021).
\bibitem{griffiths2009} J.B. Griffiths, J. Podolsky, \textit{Exact Space-Times in Einstein's General Relativity}, (Cambridge University Press, Cambridge, 2009).
\bibitem{raifeartaigh2014} C. O’Raifeartaigh et. al.,
\textit{Eur. Phys. J. H} \textbf{39}, 353 (2014).
\bibitem{eddington1930} A.S. Eddington, \textit{Mon. Not. Roy. Astron. Soc.} \textbf{90}, 668 (1930).
\bibitem{harrison1967} E.R. Harrison, \textit{Rev. Mod. Phys.}, \textit{Rev. Mod. Phys.} \textbf{39}, 862 (1967).
\bibitem{barrow2003} J.D. Barrow, G.F.R. Ellis, R. Maartens, C.G. Tsagas, \textit{Class. Quantum Grav.} \textbf{20}, L155 (2003).
\bibitem{barrow2012} J.D. Barrow, K. Yamamoto, \textit{Phys. Rev. D} \textbf{85}, 083505 (2012).
\bibitem{starobinsky1980} A. A. Starobinsky, \textit{Phys. Lett. B} \textbf{91}, 99 (1980).
\bibitem{starobinsky1983} A. A. Starobinsky, \textit{Sov. Astron. Lett.} \textbf{9}, 302 (1983).
\bibitem{canonico2010} R. Canonico, L. Parisi, \textit{Phys. Rev. D} \textbf{82}, 064005 (2010).
\bibitem{bag2014} S. Bag, V. Sahni, Y. Shtanov, \textit{J. Cosmology Astropart. Phys.} \textbf{07}, 034 (2014).
\bibitem{heydarzade2015} Y. Heydarzade, F. Darabi, \textit{J. Cosmology Astropart. Phys.} \textbf{04}, 028 (2015).
\bibitem{heydarzade2016} Y. Heydarzade, F. Darabi, K. Atazadeh, \textit{Ap\&{SS}} \textbf{361}, 250 (2016).
\bibitem{boehmer2015} C.G. Boehmer, N. Tamanini, M. Wright, \textit{Phys. Rev. D} \textbf{92}, 124067 (2015).
\bibitem{barrow1983} J.D. Barrow, A.C. Ottewill, \textit{J. Phys. A} \textbf{16}, 2757 (1983).
\bibitem{bohmer2007} C.G. Bohmer, L. Hollenstein, F.S.N. Lobo, \textit{Phys. Rev. D} \textbf{76}, 084005 (2007).
\bibitem{seahra2009} S.S. Seahra, C.G. Bohmer, \textit{Phys. Rev. D} \textbf{79}, 064009 (2009).
\bibitem{sharif20201} M. Sharif, S. Saleem, \textit{Mod. Phys. Lett. A} \textbf{35}, 2050152 (2020).
\bibitem{bohmer2010} C. G. Bohmer and F. S. N. Lobo, \textit{Eur. Phys. J. C} \textbf{70}, 1111 (2010).
\bibitem{maeda2010}   K.-I. Maeda, Y. Misonoh and T. Kobayashi,, \textit{Phys. Rev. D} \textbf{82}, 064024 (2010).
\bibitem{paul2010}  B. C. Paul and S. Ghose, \textit{Gen. Relativ. Gravit.} \textbf{42}, 795 (2010).
\bibitem{gohain2024}  M. M. Gohain, K. Bhuyan, arxiv: 2404.01355 [gr-qc].
\bibitem{paris2012}  L. Parisi, N. Radicella, and G. Vilasi, \textit{Phys. Rev. D} \textbf{86}, 024035 (2010).
\bibitem{zhang2013}   K. Zhang, P. Wu, and H. Yu, \textit{Phys. Rev. D} \textbf{87}, 063513 (2013).
\bibitem{boehmer2013}  C. G. Boehmer, F. S. N. Lobo and N. Tamanini, \textit{Phys. Rev. D} \textbf{88}, 104019 (2013).
\bibitem{atazade2014}  K. Atazadeh, \textit{J. Cosmology Astropart. Phys.} \textbf{06}, 020 (2014).
\bibitem{huang2014}  H. Huang, P. Wu and H. Yu, \textit{Phys. Rev. D} \textbf{89}, 103521 (2014).
\bibitem{shabani2017}  H. Shabani, A. H. Ziaie, \textit{Eur. Phys. J. C} \textbf{77}, 31 (2017).
\bibitem{sharif20202} M. Sharif, S. Saleem, \textit{Mod. Phys. Lett. A} \textbf{35}, 2050222 (2020).
\bibitem{shabani2019}  H. Shabani, A. H. Ziaie, \textit{Eur. Phys. J. C} \textbf{79}, 270 (2019).
\bibitem{shabani2022}  H. Shabani, A. H. Ziaie, H. Moradpour, \textit{Ann. Phys.} \textbf{444}, 169058 (2022).
\bibitem{khodadi2022} M. Khodadi, A. Allahyari, S. Capozziello, \textit{Phys. dark Univ.} \textbf{36}, 101013 (2022).
\bibitem{akarsu2023} \"{O}. Akarsu, N. M. Uzun, \textit{Phys. dark Univ.} \textbf{40}, 101194 (2023).
\bibitem{sharif2023} M. Sharif, M. Z. Gul, \textit{Universe} \textbf{9}, 145 (2023).
\bibitem{huang2023} Q. Huang, H. Huang, B. Xu, \textit{Phys. dark Univ.} \textbf{41}, 101262 (2023).
\bibitem{barca2023} G. Barca, G. Montani, A. Melchiorri, \textit{Phys. Rev. D} \textbf{108}, 063505 (2023).
\bibitem{palermo2022}O. Palermo et. al., 
\textit{Eur. Phys. J. C} \textbf{82}, 1146 (2022).
\bibitem{huang2020} Q. Huang et. al.,
\textit{Class. Quantum Grav.} \textbf{37}, 195002 (2020).
\bibitem{li2017} S. Li and H. Wei, \textit{Phys. Rev. D} \textbf{95}, 023531 (2017)
\bibitem{fadel2022}  M. Fadel and M. Maggiore, \textit{Phys. Rev. D} \textbf{105}, 106017  (2022).
\bibitem{bosso2022}  P. Bosso, L. Petruzziello, and F. Wagner \textit{Phys. Lett. B} \textbf{834}, 137415 (2022).
\bibitem{segreto2023} S. Segreto and G. Montani, \textit{Eur. Phys. J. C} \textbf{83}, 385 (2023).



\bibitem{review_fT} S. Bahamonde et. al., \textit{Rep. Prog. Phys.} \textbf{86} 026901 (2023).
\bibitem{review_fQ} L. Heisenberg, \textit{Phys. Rep.} \textbf{1066}, (2024), pages 1-78.

\bibitem{FLRW/connection} N. Dimakis et. al.,
\textit{Phys. Rev. D} \textbf{106}, 043509 (2022).
\bibitem{wang2024} Q. Wang, X. Ren, Y.-F. Cai, W. Luo, and E. N. Saridakis Observational Test of $f(Q)$ Gravity with Weak Gravitational Lensing, \textit{Astrophys. J.} 974, 7 (2024).
\bibitem{starobinsky1987} A. A. Starobinsky, A New Type of Isotropic Cosmological Models Without Singularity,
\textit{Adv. Ser. Astrophys. Cosmol.} \textbf{3}, 130 (1987)
\bibitem{lin2021} R.-H Lin and X.-H. Zhai, Spherically symmetric configuration in $f(Q)$ gravity, \textit{Phys. Rev. D} \textbf{103}, 124001  (2021).
\bibitem{araujo2024} 
J.C.N. de Araujo and H.G.M. Fortes, Compact stars in $f(Q)=Q+\xi Q^2$ gravity, [gr-qc/2407.08884].
\bibitem{pradhan2024} S. Pradhan, R. Solanki, P.K. Sahoo, Cosmological constraints on $f(Q)$ gravity models in the non-coincident formalism, \textit{Journal of High Energy Astrophysics} 43 (2024) 258. 
\bibitem{silva2024} A.M. Silva, M.J. Rebouças, N.A. Lemos, G\"{o}del-type spacetimes in $f(Q)$ gravity, [gr-qc/2407.04750].
\bibitem{fQec2} G. Subramaniam et. al., \textit{Phys. Dark Univ.} 41 (2023) 101243. 
\bibitem{nojiri2024} S. Nojiri and S.D. Odintsov, Well-defined $f(Q)$ gravity, reconstruction of FLRW spacetime and unification of inflation with dark energy epoch, \textit{Phys. Dark Univ.} 45 (2024) 101538. 
\bibitem{vishwakarma2024} P. Vishwakarma, P. Shah and K. Bamba, Autonomous system analysis of the late-time cosmological solutions and their stability in $f(Q)$ Gravity Models, [gr-qc/2401.05455]
\bibitem{rana2024} D.S. Rana, R. Solanki and P.K. Sahoo, Phase-space analysis of the viscous fluid cosmological models in the coincident $f(Q)$ gravity, \textit{Phys. Dark Univ.} 43 (2024) 101421.


\bibitem{FLRW/connection1}N. Dimakis et. al.,
\textit{Phys. Rev. D} \textbf{106}, 123516 (2022).

\bibitem{ad/viability}  A. De and T. H. Loo, 
\textit{Classical Quantum Gravity} \textbf{40} 115007 (2023). 

\bibitem{fQcosmohamid}H. Shabani et al, \textit{Eur. Phys. J. C} \textbf{84} 285 (2024).


\bibitem{palia}A. Paliathanasis, \textit{Phys. Dark Univ.}, \textbf{42} 101355 (2023).
\bibitem{jensko}E. Jensko, arXiv:2407.17568 [gr-qc].

\bibitem{coincident} 
J. B. Jim\'enez, L. Heisenberg, and T. Koivisto, \textit{Phys. Rev. D} \textbf{98}, 044048 (2018).
\bibitem{zhao} D. Zhao, \textit{Eur. Phys. J. C} \textbf{82}, 303 (2022).

\bibitem{hyper}
F. W. Hehl, G. D. Kerlick, and P. van der Heyde, 
\textit{Z. Naturforsch. A} \textbf{31}, 111  (1976).

\bibitem{de-loo-saridakis}
A. De, T. H. Loo and E. N. Saridakis, 
\textit{JCAP} \textbf{03}, 050  (2024).


\bibitem{lakhal2019} B. S. Lakhal and A. Guezmir, \textit{J. Phys.: Conf. Ser} \textbf{1269}, 012017 (2019).
\bibitem{singal2024} A. K. Singal, \textit{Eur. Phys. J. C} \textbf{84}, 385 (2024).












\end{thebibliography}
\end{document}